\documentclass[letterpaper]{article} 
\usepackage{aaai25}  
\usepackage{times}  
\usepackage{amsfonts}
\usepackage{helvet}  
\usepackage{courier}  
\usepackage[hyphens]{url}  
\usepackage{graphicx} 
\usepackage{natbib}  
\usepackage{caption} 
\usepackage{algorithm}
\usepackage{algorithmic}
\usepackage{xcolor}

\usepackage{newfloat}
\usepackage{listings}
\DeclareCaptionStyle{ruled}{labelfont=normalfont,labelsep=colon,strut=off} 
\floatstyle{ruled}
\newfloat{listing}{tb}{lst}{}
\floatname{listing}{Listing}
\title{Coordinated Inauthentic Behavior on TikTok:\\ Challenges and Opportunities for Detection in a Video-First Ecosystem}
\author{
    Luca Luceri,\equalcontrib
    Tanishq Vijay Salkar,\equalcontrib
    Ashwin Balasubramanian, \\
    Gabriela Pinto,
    Chenning Sun,
    Emilio Ferrara
}
\affiliations{
     Information Sciences Institute, University of Southern California, Marina Del Rey, California, United States\\

}

\usepackage{bibentry}

\begin{document}

\maketitle

\begin{abstract}

Detecting coordinated inauthentic behavior (CIB) is central to the study of online influence operations. However, most methods focus on text-centric platforms, leaving video-first ecosystems like TikTok largely unexplored. To address this gap, we develop and evaluate a computational framework for detecting CIB on TikTok, leveraging a network-based approach adapted to the platform’s unique content and interaction structures. Building on existing approaches, we construct user similarity networks based on shared behaviors, including synchronized posting, repeated use of similar speech segments, and multimedia content reuse, and apply graph pruning techniques to identify dense networks of likely coordinated accounts. Analyzing a dataset of 1.35M TikTok videos related to the 2024 U.S. Presidential Election, we uncover a range of coordinated activities, from synchronized amplification of political narratives and semi-automated content replication to AI-generated voiceovers and manufactured split-screen video formats. Our findings demonstrate that while several indicators proposed in this study effectively detect CIB on TikTok, other platform-native signals, such as video-based replies and Duet and Stitch interactions, often reflect organic engagement rather than inauthentic behavior, highlighting the platform’s distinct content dynamics and interaction mechanics. This work provides the first empirical foundation for studying CIB on TikTok, paving the way for future research into influence operations on video platforms. 

\textbf{Code}: \url{https://tinyurl.com/56cjkvr9}

\end{abstract}

%

\section{Introduction}
\label{sec:introduction}
With over 1.6B active users globally,\footnote{\url{https://www.statista.com/statistics/272014/global-social-networks-ranked-by-number-of-users/}}
TikTok has become one of the most influential social media platforms, playing a central role in shaping public discourse, especially among younger audiences. While its short-form video format offers an engaging space for creativity and self-expression, it has also become a vector for influence campaigns aiming to sway public opinion, disseminate disinformation, and manipulate political narratives. Such campaigns are particularly concerning during high-stakes events like national elections or public health crises, where coordinated dissemination of false or polarizing content can exacerbate societal divisions, trigger panic, or erode trust in institutions. Recent internal audits and journalistic investigations have highlighted TikTok’s growing role in covert influence operations,\footnote{\url{https://www.tiktok.com/transparency/en/covert-influence-operations?enter_method=category_card}}
\footnote{\url{https://www.cnn.com/2024/09/24/tech/tiktok-russian-media-accounts-us-election}}
\footnote{\url{https://www.nbcnews.com/tech/tech-news/tiktok-says-disrupted-15-influence-operations-year-one-china-rcna153831}}
underlining the urgent need for systematic study and effective monitoring of the platform’s information dynamics.

Coordinated inauthentic behavior (CIB) is widely recognized as a hallmark tactic in online influence campaigns, where multiple accounts, often automated or inauthentic, amplify narratives in a synchronized fashion~\cite{pacheco2021uncovering,luceri2024unmasking,tardelli2024multifaceted}. These accounts often post similar content in close temporal proximity, reuse media assets, or repeatedly interact with one another to boost engagement metrics. While not inherently malicious, such coordination becomes problematic when it is deceptive in nature, e.g., when inauthentic accounts present themselves as genuine individuals or manipulate visibility through algorithmic gaming. These tactics are commonly used to promote specific political agendas, manufacture the illusion of consensus, or suppress opposing viewpoints, particularly during moments of societal stress
\cite{vishnuprasad2024tracking,alizadeh2020content,cinus2025exposing}.


Research on detecting CIB has largely focused on Twitter/$\mathbb{X}$ \cite{magelinski2022synchronized,Pacheco_2020,tardelli2024temporal,luceri2024leveraging}, where behavioral traces like co-retweets, shared URLs, and posting synchronicity have proven useful for uncovering coordinated efforts like state-linked information operations. A smaller body of work has investigated coordination on other platforms, such as Facebook and Telegram \cite{giglietto2020detecting,giglietto2020takes,cinus2025exposing,ng2024tiny}. 
In contrast, video-first platforms like TikTok remain understudied, 
with no prior work systematically examining CIB on the platform using computational methods.

This gap is particularly important given the algorithmic design of TikTok’s \textit{For You} feed, which personalizes content based on behavioral signals like viewing time, likes, and shares, rather than social connections or follower graphs. This design amplifies the reach of content regardless of follower count and makes it easier for coordinated campaigns to achieve high visibility if they trigger the right interaction patterns \cite{bandy2020tulsaflop}. As a result, a coordinated set of accounts can potentially manipulate visibility by posting similar content or interacting in patterns that exploit the algorithm’s engagement-based logic. Unlike Twitter/$\mathbb{X}$, where retweet networks and reply chains offer structural clues about coordination, TikTok lacks explicit interaction graphs, and reposting is relatively rare. Moreover, TikTok’s dominant use of video and audio complicates the direct application of text-based similarity methods that have proven effective on text-dominant platforms like Twitter/$\mathbb{X}$, Facebook, and Telegram. As a result, the detection of coordination on TikTok requires methodological adaptations tailored to its media format and engagement mechanics.

\paragraph{Contribution of this Work.}
To address this gap, we develop and evaluate a computational framework for detecting CIB on TikTok, leveraging a network-based approach adapted to the platform’s unique content and interaction structures. Building on established methodologies developed for Twitter/$\mathbb{X}$ \cite{luceri2024unmasking,pacheco2021uncovering,cinus2025exposing}, we first construct similarity networks that link users based on shared behaviors, such as uploading videos within tightly bounded time windows, or repeating similar audio and video content. Then, we apply graph pruning techniques to isolate dense subnetworks of potentially coordinated inauthentic accounts. Leveraging this approach, we analyze a dataset of 1.35M TikTok videos related to the 2024 U.S. Election \cite{pinto2025tracking} to explore how inauthentic coordination manifests in the TikTok ecosystem. 

Our findings highlight both promising opportunities and key challenges in building effective CIB detection strategies tailored to TikTok’s unique dynamics. Several behavioral signals—some newly introduced in this study, others adapted from prior work on Twitter/$\mathbb{X}$—proved highly effective in surfacing likely coordination clusters. These clusters exhibited patterns consistent with campaign-style amplification, including synchronized dissemination of political content, repeated posting of videos with visually identical formats and AI-generated voiceovers, and the use of split-screen formats to replicate messaging across accounts while circumventing content moderation. In many cases, accounts shared identical voice tracks and promoted the same external domains, including monetization platforms and mock websites—a commonly observed pattern on other platforms \cite{minici2024uncovering}.\footnote{\url{https://www.veraai.eu/posts/uni-urbino-on-cib-around-pope-health-research-findings}}
Additional signals of inauthenticity included auto-generated or extremely similar usernames, frequent handle alterations and swapping, and high-volume posting of near-duplicate content, suggesting semi-automated or orchestrated account control.

However, not all indicators yielded meaningful results. TikTok-specific features, such as video-based reply, duet, and stitch, did not produce clusters indicative of coordination. These results likely reflect the platform’s unique norms of content creation and interaction, where reuse and remixing are common and often organically driven. 
While these TikTok-specific features did not yield strong evidence of CIB, 
our framework uncovered novel signatures of coordination unique to TikTok, such as watermark reuse, synthetic voice patterns, including audio deep fake, and the strategic use of manufactured split-screen videos. These findings suggest that TikTok-native signals hold significant potential for future detection efforts and should be incorporated into computational models designed to capture the evolving nature and increasingly sophisticated, AI-driven dynamics of inauthentic behaviors on video-first platforms.

Despite mounting evidence of coordinated manipulation, platform enforcement appears limited: Only a small fraction of the suspicious accounts identified through our analysis were suspended or removed. This gap underscores the urgent need for proactive detection and moderation strategies. By systematically evaluating a diverse range of behavioral signals on a large-scale, election-related TikTok dataset, this study provides the first empirical foundation for detecting coordination on the platform and offers concrete directions for future research in this emerging area.

\section{Related Work}

\subsection{Influence and Misinformation Modeling on TikTok} 

Understanding how influence operates on social media platforms is critical for analyzing content diffusion and coordinated behavior. Classical models of influence and information diffusion underscore the role of repeated exposure in belief propagation \cite{centola2010spread}. On TikTok, influence is primarily mediated by algorithmic recommendation systems that amplify user preferences through short-form videos, trending hashtags, and viral visual and sound effects. Recent research shows that TikTok's \textit{For You} page can foster echo chambers by reinforcing homophily and interaction-based feedback loops \cite{zeng2022content,gao2023echo}. These exposure-driven dynamics are especially potent when political or controversial content is involved, often resulting in the rapid spread of low-credibility information, even with minimal exposure \cite{ye2024susceptibility}.

Misinformation on TikTok, particularly in multimodal formats, presents additional challenges \cite{shang2025multitec}. 
\citet{corso2024conspiracy} examined the spread of conspiracy theories on TikTok. Their large-scale analysis identified a lower-bound prevalence of approximately 0.1\% for conspiracy-laden videos, revealing how even a small fraction of such content can leverage TikTok’s algorithmic infrastructure to achieve wide reach. The study underscores the unique challenges posed by conspiratorial narratives in short-form, remixable formats, which often evade detection and contribute to the broader misinformation ecosystem.

Similarly, \citet{zenone2022using} analyzed TikTok videos related to the 2022 monkeypox outbreak and identified videos propagating several conspiracy theories. Their findings highlight the importance of social media monitoring as a tool for detecting and addressing public health misinformation. 
Notably, TikTok creators have reported frustration with the insufficient moderation of harmful comments and misleading content, raising further concerns about the platform's role in content governance \cite{zeng2022content}. To address this issue, \citet{shang2021multimodal} propose \textit{TikTec}, a framework designed to detect COVID-19 misinformation on TikTok. Their findings emphasize the importance of integrating visual features, captions, and speech signals to improve detection performance.

\subsection{Detection of Coordinated Inauthentic Behavior}
CIB has emerged as a key behavioral pattern underpinning the spread of disinformation, propaganda, and conspiratorial narratives across social media platforms. While coordination is not inherently malicious, its deceptive forms are typically characterized by manipulative amplification of highly partisan or misleading content \cite{nizzoli2021coordinated}.

The detection of such orchestrated efforts has evolved from early machine learning approaches, which primarily focused on distinguishing automated accounts (i.e., bots) from genuine users \cite{Yang_2019, cresci2016dna}, to more advanced models capable of identifying human-operated inauthentic actors, including state-sponsored trolls. Recent studies have employed content-based features \cite{alizadeh2020content, luceri2024leveraging}, behavioral signatures \cite{luceri2020detecting, sharma2022characterizing}, and temporal or sequence-based patterns \cite{ezzeddine2022characterizing, nwala2023language} to capture covert coordination strategies.

Among the most effective paradigms are network-based coordination detection techniques, which model user behaviors as similarity graphs constructed from shared hashtags, URLs, posting times, or interaction targets \cite{Pacheco_2020, pacheco2021uncovering, magelinski2022synchronized, nizzoli2021coordinated}. These similarity networks enable the application of topological measures, such as edge weights capturing behavioral similarity and centrality metrics to identify influential or densely connected nodes, to isolate clusters of users potentially involved in influence operations \cite{luceri2024unmasking,minici2025iohunter, mannocci2024detection, tardelli2024temporal, vishnuprasad2024tracking}. Much of this literature has focused on Twitter, leveraging its structured interaction data (e.g., retweets) to surface coordination. Parallel work has explored intra- and inter-platform CIB dynamics on different social media networks, such as Twitter, Facebook, Telegram, YouTube, and 4chan, also examining how content and actors migrate between ecosystems \cite{cinus2025exposing,ng2022cross, zannettou2019disinformation, gatta2023interconnected}. 


In contrast, our work introduces a fine-grained, unsupervised CIB detection framework for TikTok, a platform that has thus far received little attention in this domain. By inspecting traditional, novel, and TikTok-specific behavioral indicators, this paper represents the first systematic attempt to extend network-based CIB detection to TikTok, providing both methodological innovations and empirical insights into coordinated influence on short-form video platforms.

\section{Methodology}

\subsection{Data Curation, Collection, and Processing}\label{sec:data}
This study leverages the public TikTok 2024 U.S. Presidential Election Dataset released by \citet{pinto2025tracking}, a large-scale, multimodal collection capturing TikTok video content shared in the lead-up to the 2024 U.S. Presidential Election. The dataset was assembled using the TikTok Research API, with videos filtered through a curated and evolving set of election-related hashtags and keywords (listed in the \textit{Appendix}) aligned with trending discourse throughout the collection period. The full dataset spans the period from November 1, 2023, to January 20, 2025.


For this study, we analyze a narrower subset of videos published between August 1 and October 31, 2024, a critical pre-election window marked by intensified campaign activity and heightened potential for coordinated messaging. This subset comprises approximately 1.35M videos posted by 362K unique users, capturing key events such as the official nomination of Kamala Harris and the first presidential debate. Additionally, we retrieved video files and associated comments using a third-party crawler. We were able to successfully download comments for approximately 92\% of the videos; some retrievals failed due to videos being deleted, made private, or shared by suspended accounts.








\subsection{Methodological Framework for CIB Detection}\label{sec:framework}
Our coordination detection framework builds on network-based methodologies established in prior research \cite{pacheco2021uncovering,luceri2024unmasking,cinus2025exposing}, and is adapted to reflect the unique content modalities and interaction dynamics of the TikTok platform. 

 The core assumption underlying this approach is that anomalously high behavioral similarity between users, across content, timing, or interaction patterns, can indicate underlying coordination, particularly when such overlap is unlikely to occur organically (e.g., due to exogenous factors, such as major external events, that could steer collective behavior). Accordingly, our pipeline detects such patterns through three main stages:
(1) the extraction of behavioral traces that may indicate coordination;
(2) the construction of similarity networks for each behavioral signal; and
(3) the detection of CIB clusters through graph pruning techniques.

\subsubsection{(1) Extracting Behavioral Traces.}\label{sec:behavioral_traces}
We define a set of behavioral traces that capture different dimensions of user similarity on TikTok. While some of these traces parallel those used on platforms like Twitter/$\mathbb{X}$, others are specifically introduced to account for TikTok’s platform-specific modalities and interaction norms. These behavioral signals, extracted from video content, descriptions, metadata, and comments, form the basis for constructing user similarity graphs across various coordination signals:

\begin{itemize}
\item \textit{Hashtag Sequence:} 
Users posting videos with 
the same set of hashtags in the same exact order,
especially when repeated across multiple accounts, can signal coordinated attempts to influence discoverability and visibility in TikTok's algorithmically curated feed.

\item \textit{Synchronized Posting:} Users who post content within narrow time windows, typically a few minutes apart,
may indicate that multiple accounts are operating in a synchronized manner, through automation or centralized control, to maximize visibility and impact by flooding the platform with similar content at the same time.

\item \textit{Co-Domain:} Users who repeatedly link to the same external domains in video descriptions (or comments) may be exhibiting coordinated behavior, especially when those domains are affiliated with low-credibility, campaign-related, or monetized websites.
To account for superficial variations (e.g., URL parameters), we extract and normalize the domain-level information (e.g., stripping query strings or subpaths), allowing us to detect shared linking behavior even when full URLs differ.

\item \textit{Co-Duet and Co-Stitch:} TikTok allows users to combine their videos with others using Duet or Stitch. In a \textit{Stitch}, the video from another user is played first, followed by the user's own content. In a \textit{Duet}, the screen is split into two parts, showing the original and user-generated videos side by side. While this feature is commonly used to engage with content from celebrities, media outlets, and viral creators, it may be used for coordinated attacks. 

\item \textit{Co-Reply:} TikTok allows users to respond to comments through video replies. These replies display the original comment as a text overlay and are marked by descriptions like “Replying to @username.” While often used for organic engagement, this feature can potentially be exploited for coordinated activity, e.g., by repeatedly replying to comments from associated accounts to boost content visibility and engagement.

\item \textit{Speech Similarity:} Users posting videos with highly similar or identical speech segments may indicate orchestrated messaging or script-based content generation. To enable analysis of spoken language in videos, we first transcribe the audio using OpenAI’s speech-to-text model, \textit{Whisper}. High speech similarity between users, particularly when occurring repeatedly, may indicate centrally coordinated messaging strategies.

\item \textit{{Video Similarity:}} Users who post visually or structurally similar videos may be participating in orchestrated content dissemination. We leverage visual models to identify instances of near-duplicate or templated content. High video similarity across multiple accounts, especially when observed repeatedly and within narrow temporal windows, is a strong indicator of coordinated amplification. This is particularly relevant on TikTok, where platform-native video editing tools (e.g., templates, split-screens, and overlays) can facilitate subtle forms of content reuse and signal campaign orchestration.


\end{itemize}

While other indicators, such as co-reposts (i.e., multiple users re-sharing the same video) and fast reposts (i.e., re-sharing at high speed), could potentially signal coordination \cite{pacheco2021uncovering,vishnuprasad2024tracking}, the low prevalence of observable repost activity in our dataset, combined with limited accessibility through the TikTok API, prevented us from investigating these signals further.

\subsubsection{(2) Constructing Similarity Networks.}\label{sec:similarity_networks}
We construct distinct similarity networks, each corresponding to one of the seven behavioral traces examined in this study.
For each behavioral trace, except Speech and Video Similarity, we construct a bipartite graph where nodes represent users on one side and behavioral entities on the other (e.g., hashtag sequences, domains, or time bins). An edge exists between a user and an entity if the user engages with that entity. 
For instance, a user is connected to a specific hashtag sequence if they post a video with that exact ordered set of hashtags in the description, or a user is linked to a temporal bin (e.g., a 5-minute interval) if they post a video within that time window. For co-duet, co-stitch, and co-reply a user is linked to another user (as an entity) if they perform a duet, stitch, or reply interaction with their content. 

To weight edges in each bipartite graph, we apply term frequency–inverse document frequency (TF-IDF), following prior work \cite{pacheco2021uncovering,Pacheco_2020,luceri2024unmasking,cinus2025exposing}. This reduces the influence of commonly used entities, such as viral hashtags or major news domains, and elevates less frequent, potentially coordinated signals. Each bipartite graph is then projected onto the user space to form a user-user similarity network, where users are represented as TF-IDF vectors and pairwise cosine similarity is computed to estimate user similarities. The result is a weighted, undirected graph in which edge weights encode the degree of behavioral similarity between users for a given trace.

Speech Similarity and Video Similarity traces depart from the bipartite graph approach. Instead, we compute direct pairwise user similarity based on content embeddings derived from audio and visual inputs.

For the Speech Similarity behavioral trace, we compare textual content from the audio transcripts extracted with \textit{Whisper}.
All textual inputs undergo a standardized cleaning process, including the removal of punctuation and stopwords. We retain only speech segments with a minimum length of four words, as shorter inputs tend to introduce noise and yield unreliable similarity scores. To compute embeddings, we use the multilingual sentence-transformer model \textit{stsb-xlm-r-multilingual} from Hugging Face. Cosine similarity is then calculated using the FAISS library for scalable approximate nearest-neighbor search.
To isolate only the most synchronized and potentially inauthentic behaviors, we focus on user pairs exhibiting perfect similarity (cosine similarity = 1) and simultaneous posting (0-second difference), indicating both semantic and temporal alignment. To further reduce the likelihood of coincidental similarity, we require that such exact matches occur at least twice between the same user pair, consistent with prior work \cite{vishnuprasad2024tracking, Pacheco_2020}. This conservative approach emphasizes repeated, deliberate coordination rather than one-off coincidences. A link is established between two users if they satisfy this criterion, resulting in a similarity network where edges connect users who have posted highly similar content at the exact same time on multiple occasions. 

For the Video Similarity behavioral trace, we employed the ViSiL similarity model \cite{kordopatis2019visil} to generate embeddings for all videos in the dataset. These embeddings were indexed using FAISS to enable efficient pairwise similarity computation at scale. Following a procedure analogous to the Speech Similarity trace, we constructed a video similarity network, where edges connect users who repeatedly posted near-duplicate videos in a highly synchronized manner. Specifically, we retained only video pairs with a similarity score exceeding 0.9 and a posting time gap of exactly 0 seconds. An edge between two users was created only if they shared at least two such highly similar videos, ensuring that the detected patterns reflect repeated and temporally aligned video reuse.

\subsubsection{(3) Pruning Similarity Networks.} To detect clusters of potentially coordinated accounts, we prune each similarity network using two complementary strategies. In both cases, the underlying intuition is that CIB-driven campaigns, which involve multiple accounts acting in coordination, tend to exhibit a high degree of \textit{collective similarity}. In a similarity network, this is reflected by nodes that are connected to many others—indicating users whose behavior closely aligns with multiple peers. Based on this rationale, we evaluate the following two pruning approaches:

\begin{itemize}
\item \textit{Node pruning via centrality}: We compute eigenvector centrality and retain only nodes above a specified percentile, following prior work suggesting that coordinated actors tend to occupy central positions in similarity networks \cite{luceri2024unmasking}, i.e., they typically share a greater number of behavioral similarities with other users compared to organic, legitimate accounts.

\item \textit{Combination of edge filtering and node pruning}: We first remove edges below a cosine similarity threshold, and then apply the centrality-based node pruning. This combined method has been shown to be highly precise and effective in minimizing false positives \cite{cinus2025exposing}.
\end{itemize}

We apply both pruning strategies across all behavioral indicators and report the most significant and interpretable coordination patterns uncovered. While both approaches are deliberately conservative, using high percentile thresholds consistent with prior work \cite{minici2024uncovering,pacheco2021uncovering}, the second strategy, which combines edge filtering with node pruning, constitutes an even stricter method specifically designed to minimize false positives (i.e., organic, legitimate users misclassified as coordinated inauthentic accounts). This combined strategy prioritizes precision, whereas node pruning in isolation is expected to yield higher recall. We report results from the combined method in cases where node pruning alone does not surface compelling evidence of CIB. Given the novelty of our approach and its first application to TikTok, we adopt this cautious strategy to minimize the risk of misclassification and ensure high precision in detecting CIB networks.

Following pruning, we analyze the resulting network structures to assess whether clusters exhibit strong signals of CIB. We adopt a mixed-methods approach, combining automated analysis with manual inspection, and examine a wide range of features, e.g., repeated usernames, shared bios and profile images, and the co-occurrence of multiple coordination signals, such as synchronized posting, content reuse across modalities (e.g., hashtags, audio, visuals), and templated formats (e.g., split-screen or AI-generated voiceovers). Importantly, the co-occurrence of these behavioral and identity-level cues increases confidence that a given cluster is not the result of organic alignment but rather reflects likely coordinated inauthentic behavior.
Manual inspection was performed independently by two annotators, consistently yielding exceptionally high agreement (see the \textit{Appendix} for additional details).

\section{Results}\label{sec:results}

We apply the CIB detection framework separately for each behavioral trace and for each month, and summarize our findings as follows, capturing the varying effectiveness of different coordination signals in surfacing CIB.

\subsection{CIB Detection via Hashtag Sequence}
\begin{figure}[t!]
\centering
\includegraphics[trim=0cm 3cm 0cm 3cm, clip=true,width=.8\columnwidth]{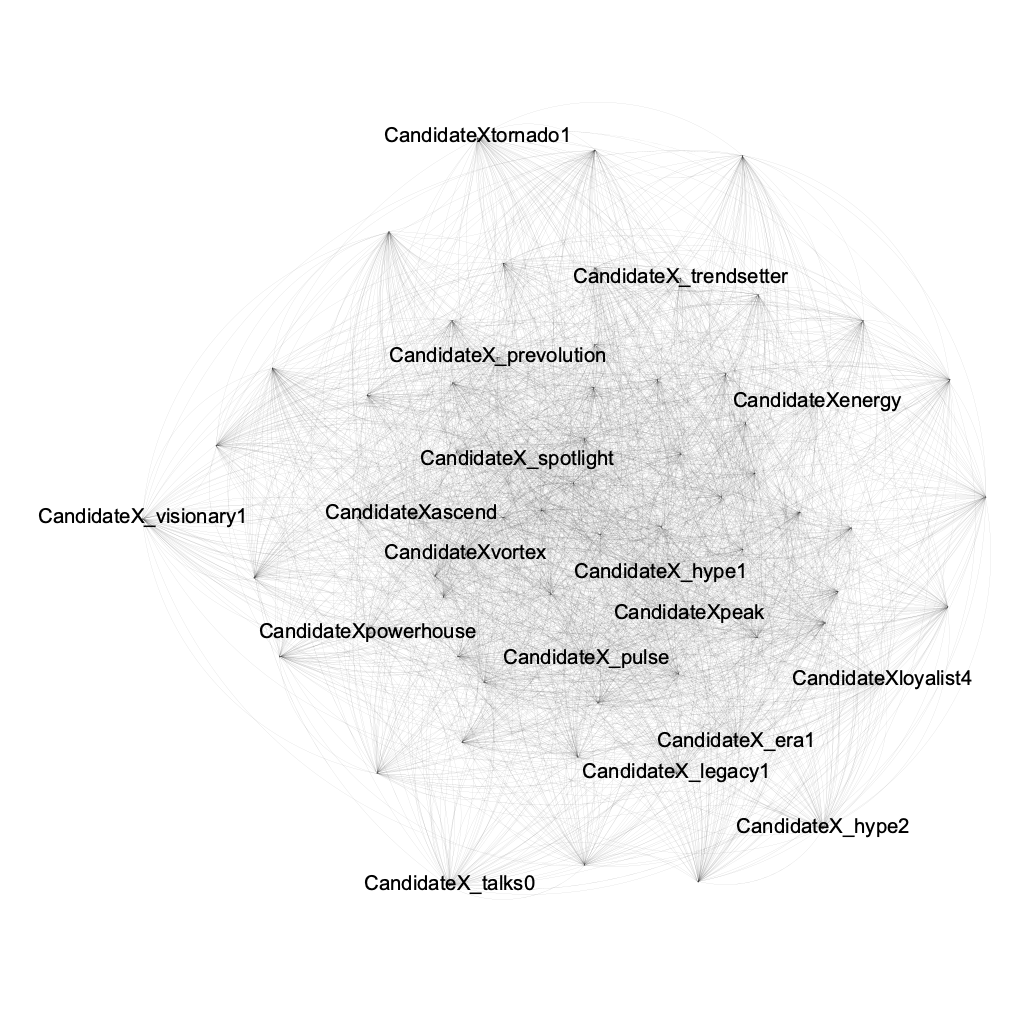} 
\caption{Cluster of coordinated accounts sharing identical hashtag sequences. Usernames within the cluster show strong similarity, all beginning with a prefix linked to a U.S. presidential candidate, denoted here as ``CandidateX" to preserve anonymity and privacy. To reduce visual clutter, we display only the usernames of accounts that posted more than five videos with AI-generated voiceovers.}

\label{fig:co-hashtag_august}
\end{figure}

We compute eigenvector centrality for each node in the \textit{Hashtag Sequence} similarity network and apply a threshold at the 98th percentile to isolate the most central and potentially coordinated users, in line with prior work \cite{minici2024uncovering,luceri2024unmasking}. Applying this filtering to user activity in August yields a dense, fully connected subgraph of 68 users (displayed in Figure~\ref{fig:co-hashtag_august}), with each node connected to all others by edges of weight 1, indicating that these accounts shared identical hashtag sequences. Such uniformity reflects a high degree of \textit{collective similarity}, a pattern associated with CIB as demonstrated by \cite{luceri2024unmasking}. This coordination is likely intended to manipulate TikTok’s recommendation algorithm through the artificial amplification of specific narratives through hashtags.

Manual inspection further revealed that all of these accounts shared highly similar usernames, each beginning with the same prefix (see Figure~\ref{fig:co-hashtag_august}), pointing to possible centralized control of these coordinated accounts. The prefix corresponds to the name of a U.S. presidential candidate. To preserve anonymity and privacy, we refer to this candidate as ``CandidateX". Examples of usernames include  ``CandidateXtornado1" and ``CandidateXgreatness."

\begin{figure}[t!]
\centering
\includegraphics[width=1\columnwidth]{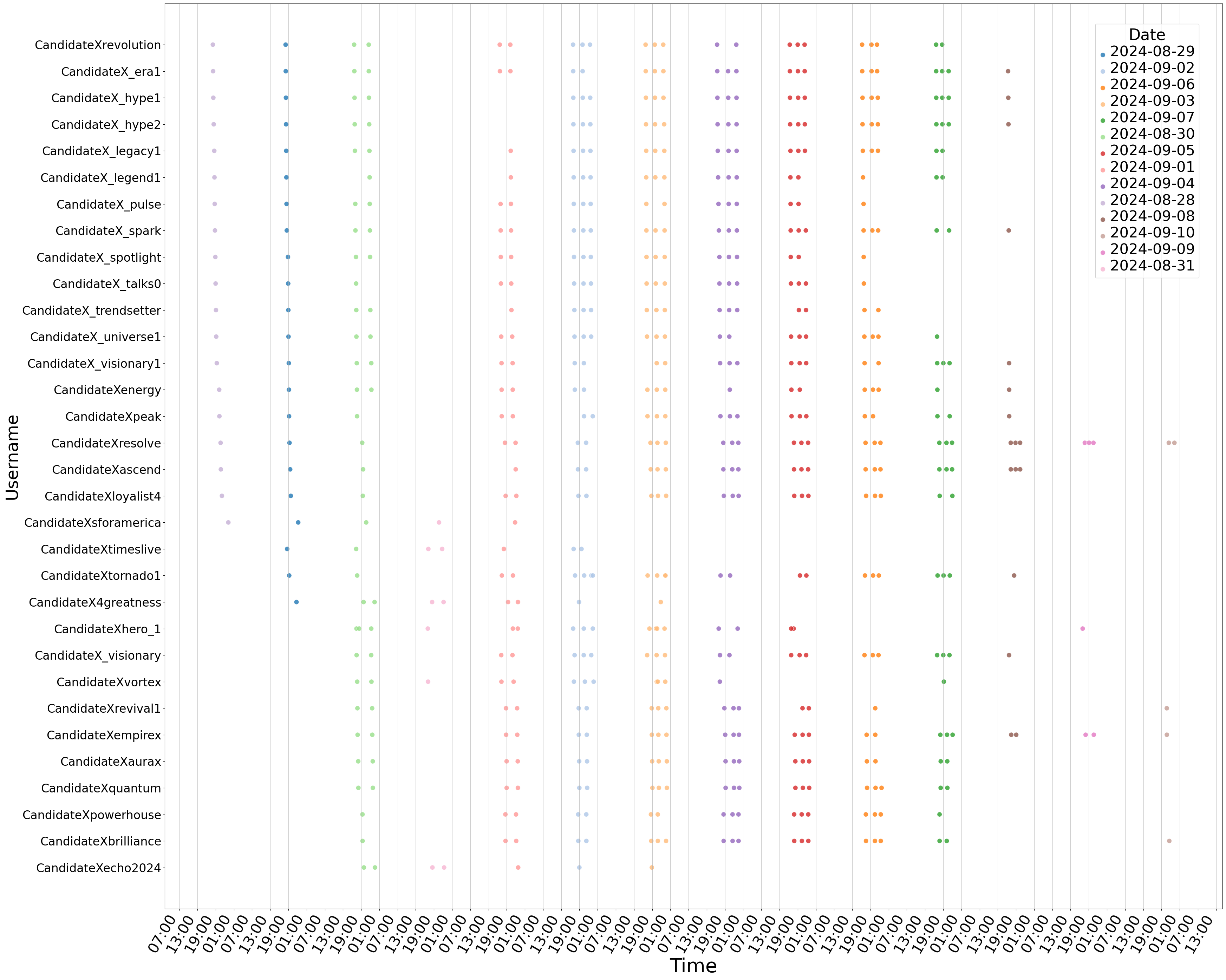} 
\caption{Synchronous posting times of coordinated accounts sharing the same hashtag sequences.}
\label{fig:sync}
\end{figure}

To further investigate CIB, we examined the posting activity of the coordinated accounts. All of them used the same hashtag sequences. The most shared sequence was: \#america, \#CandidateXNameLastName, \#CandidateXLastName, \#2024, \#election2024, appearing in 624 video descriptions. The users consistently posted one after another, with inter-post intervals of less than five minutes, indicating tightly synchronized activity. The temporal proximity of their posts over an extended period, as depicted in Figure \ref{fig:sync}, offers another strong evidence of CIB aimed to manipulate TikTok's feed algorithm through synchronized posting.

Manual review of the videos revealed an additional sign of coordination: all videos contained an identical watermark, in the top-left corner, reading ``Team CandidateX'', indicating a shared content source and reinforcing the likelihood of a centrally organized agenda.
\begin{figure}[t!]
\centering
\includegraphics[width=0.8\columnwidth]{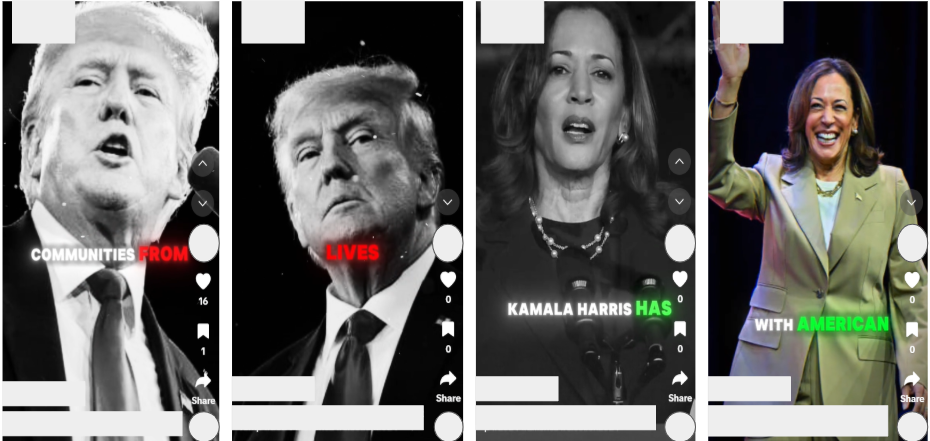} 
\caption{Examples of videos shared by coordinated accounts sharing the same hashtag sequences. The watermark in the top-left corner and the description at the bottom were obscured to preserve anonymity and privacy.}
\label{fig:videos_cohashtag}
\end{figure}
Beyond this similarity, we observed widespread duplication of both video and audio, with most videos featuring caption voiceovers. Identical videos with the same audio were often reposted across coordinated accounts, while in other cases, visuals were reused with slight audio variations (or vice versa) suggesting deliberate strategies for mass dissemination and evasion of platform detection.
Figure \ref{fig:videos_cohashtag} shows an example of identical posts shared by multiple users within the coordinated cluster.

\begin{figure}[t]
\centering
\includegraphics[width=0.85\columnwidth]{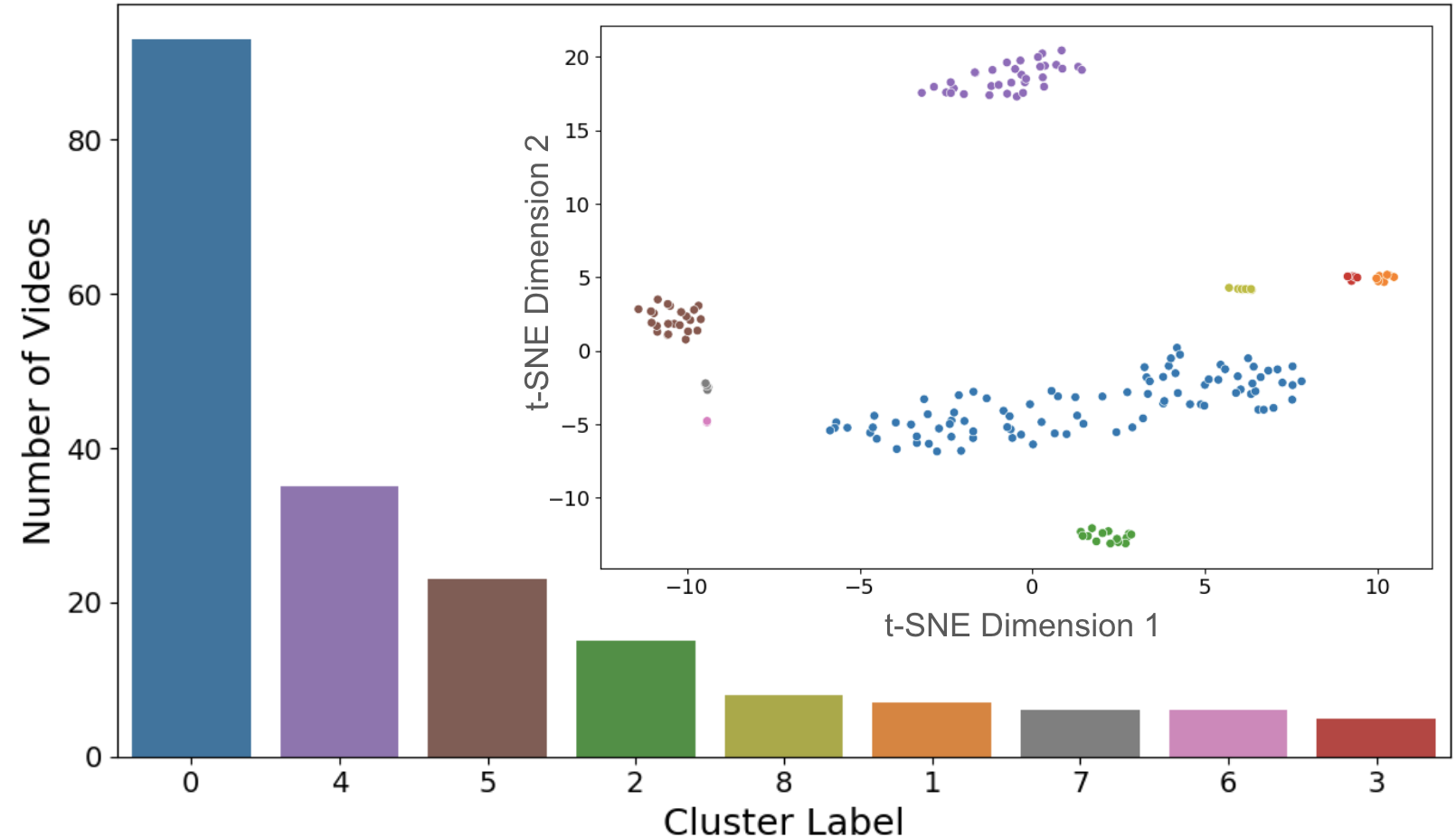} 
\caption{Clustering of spectrogram vectors extracted from videos shared by coordinated accounts using identical hashtag sequences. The inset shows a t-SNE dimensionality reduction of the spectrograms, highlighting distinct clusters corresponding to different AI-generated synthetic voices.}
\label{fig:spectrogram}
\vspace{-0.5cm}
\end{figure}

We also detected the use of AI-generated voiceovers, with over one-third of the videos featuring synthetic voices. To confirm this, we extracted audio from each video using the \textit{FFmpeg} library and computed mel-spectrograms with the \textit{librosa} library. We applied DBSCAN clustering to the spectrogram vectors, with hyperparameters selected via the k-distance graph, following the voice deepfake detection methodology described in \cite{barnekow2021creation}. As shown in Figure~\ref{fig:spectrogram}, nine distinct clusters naturally emerge. Upon inspecting each cluster, we found that all videos within a given cluster shared the same type of synthetic voice. This suggests that users in this coordinated network repeatedly use a variety of AI tools to generate voiceovers, likely in an effort to evade detection by diversifying their activity. Interested readers can refer to our code repository to explore the diverse voiceovers used in the analyzed videos.

\begin{figure}[t]
\centering
\includegraphics[width=0.5125\columnwidth]{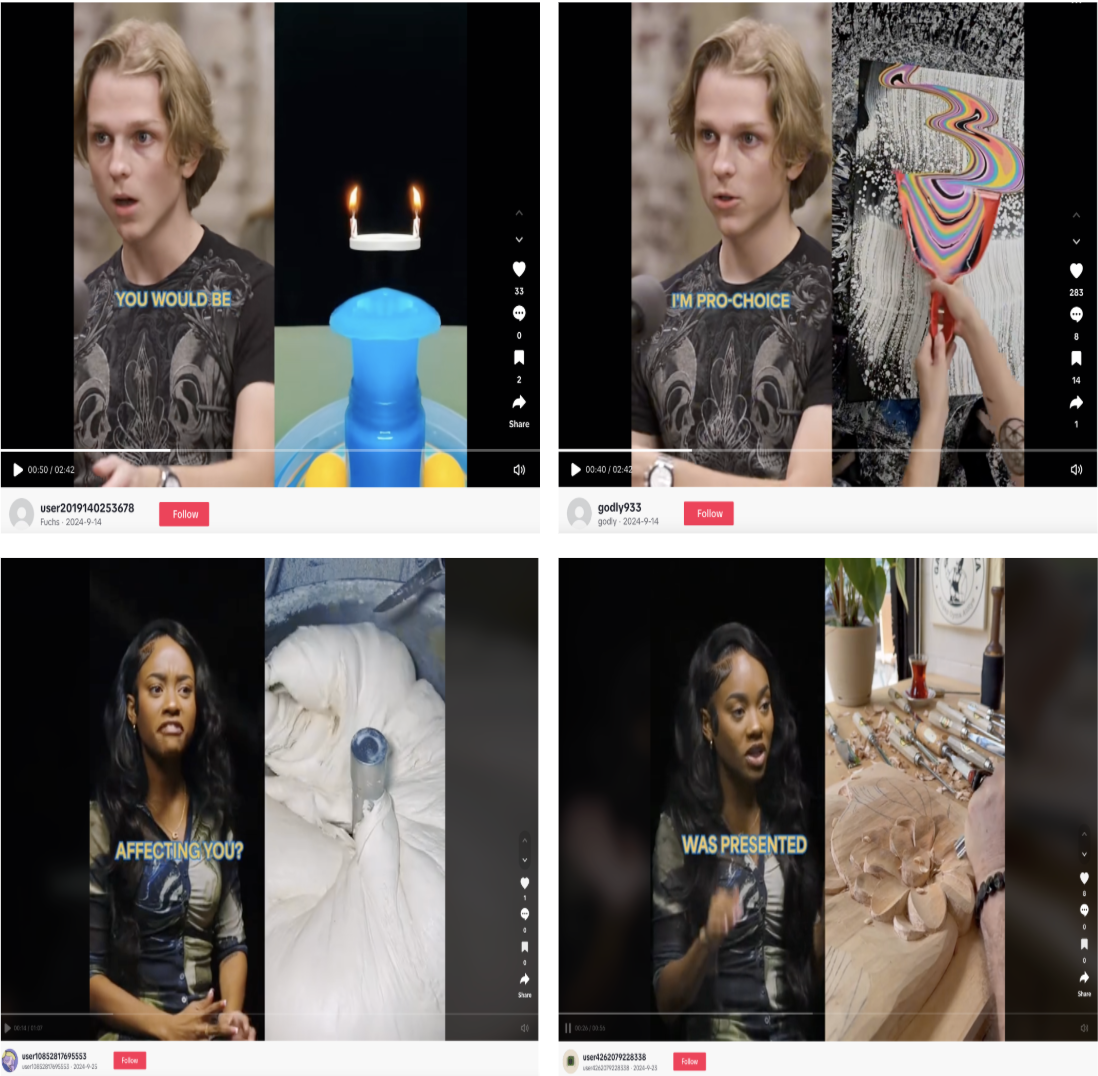} 
\includegraphics[width=0.47\columnwidth]{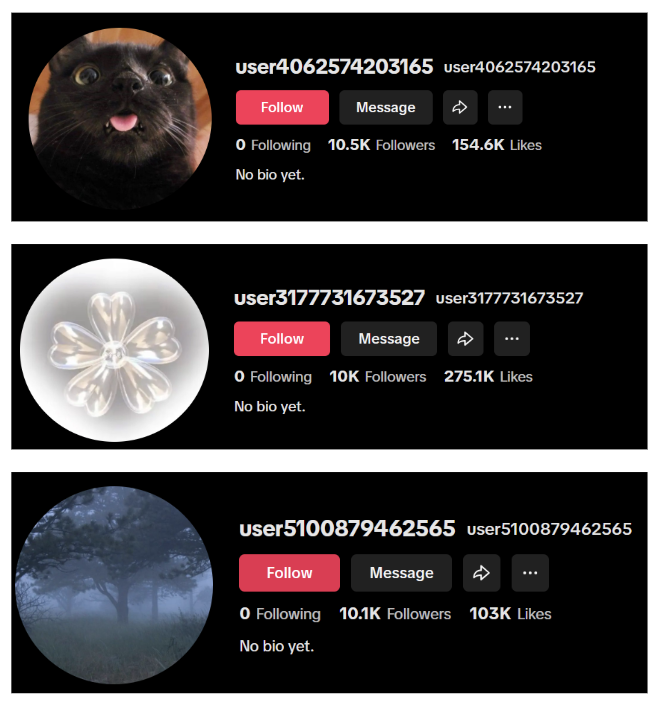} 
\caption{Examples of split-screen videos and profiles from coordinated accounts sharing identical hashtag sequences.}
\label{fig:split}
\vspace{-0.5cm}
\end{figure}

Focusing on user activity in September, we applied the same approach to the corresponding Hashtag Sequence similarity network, yielding a dense network of 68 users (a completely different set from the one found in August) with a graph density of 0.98—another clear example of \textit{collective similarity}. In addition to repeatedly sharing the same hashtag sequences, users in this cluster also posted identical video content and descriptions. Notably, they employed a \textit{split-screen} format, in which the left side of the video featured the actual content, typically political debate footage, while the right side showed unrelated visual material, such as chocolate-making (see Figure \ref{fig:split}). While they resemble a \textit{duet}, these videos are independently created and structured to imitate the format, possibly as an attempt to evade detection by obscuring the duplication of core content.

A case in point involves a popular YouTube video that was segmented and repurposed, with each segment paired with unrelated visuals to mask repetition. 
The content itself was highly polarized, addressing themes related to the 2024 U.S. Presidential Election and sensitive sociopolitical topics, including abortion, race, gender identity, and religion.


Finally, we observed that 21 of the 68 accounts had auto-generated usernames, consisting of the prefix ``user" followed by a long numeric string (12–14 digits), e.g., @user1234567890000. This deviates significantly from typical user naming patterns and provides further evidence of possible automated or inauthentic account creation. This is further supported by profile-level inspection of these accounts (see Figure \ref{fig:split}), which often feature random or generic profile pictures, empty bios, and inflated follower counts, suggesting the possible use of fake or purchased followers to simulate legitimacy and reach.

In October, we identified a coordinated network of 19 users, now all removed or suspended from TikTok, who engaged in strikingly synchronized behavior. These accounts posted 157 videos featuring identical hashtag sequences and nearly identical descriptions that included ``CandidateX'' name and a numeric identifier that varied systematically between posts, suggesting the use of a templated posting strategy. Activity peaked on October 17th and 18th, with many videos published within tight time windows. Taken together, the combination of identical hashtags, near-duplicate descriptions, and synchronized posting provides strong evidence of coordinated inauthentic behavior.


\subsection{CIB Detection via co-Domain}

Similar to the Hashtag Sequence behavioral trace, we
pruned the co-Domain similarity network using an eigenvector centrality threshold at the 98th percentile. For the month of August, this process yielded a fully connected cluster of 16 users. All usernames in this cluster either began with or contained the string ``CandidateY'', suggesting alignment with or support for a specific presidential candidate. As in previous examples, we use ``CandidateY'' as a placeholder to maintain anonymity and privacy.

\begin{figure}[t]
\centering
\includegraphics[trim=4.5cm 0cm 4cm 3cm, clip=true, width=0.77\columnwidth]{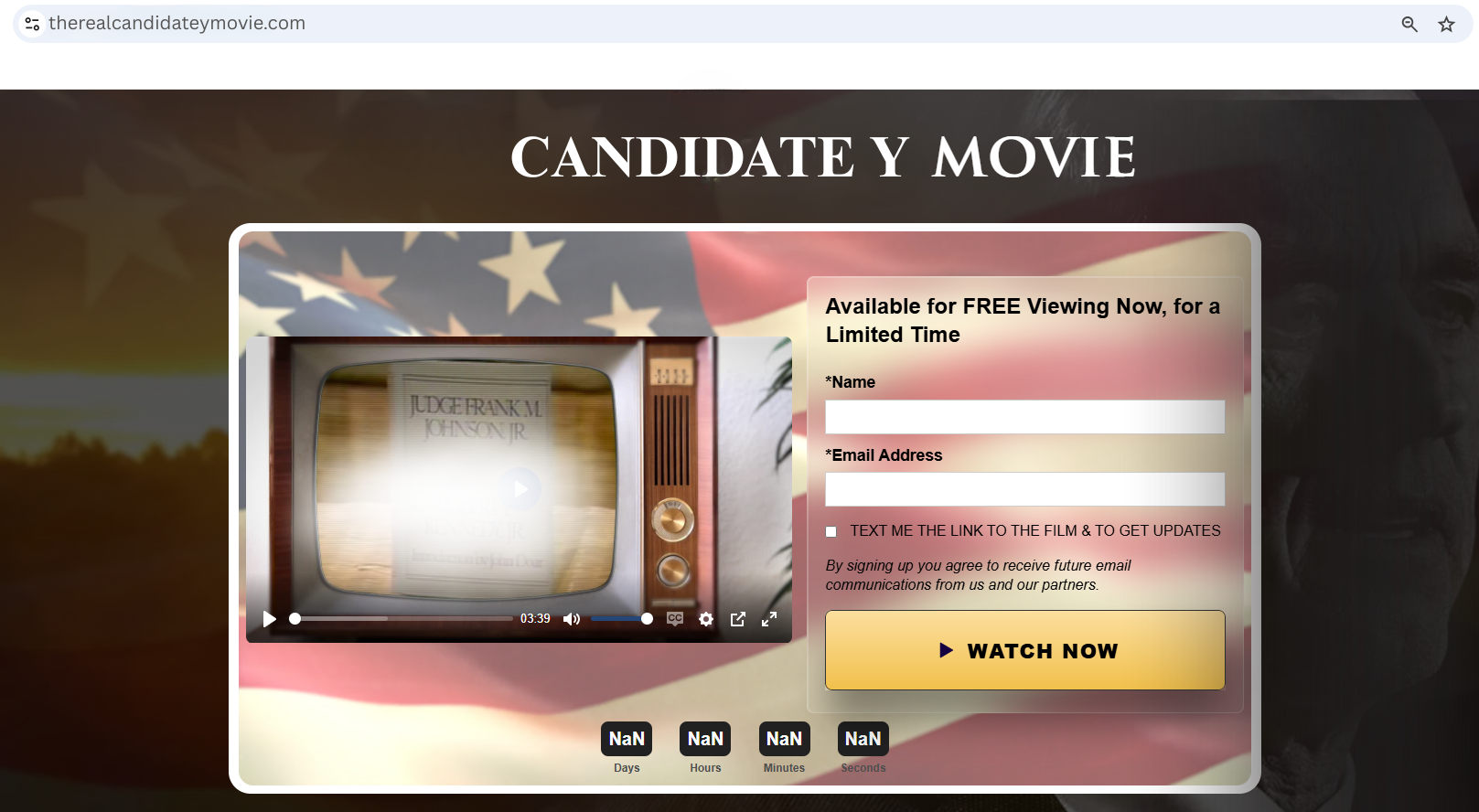} 
\caption{Web page promoted by coordinated accounts identified through the co-domain indicator.}
\label{fig:co-domain_august}
\vspace{-0.5cm}
\end{figure}

All of these accounts included URLs in their video descriptions that redirected to the same website, shown in Figure~\ref{fig:co-domain_august} with candidate-specific details intentionally obscured for anonymity and privacy. Although the web domains varied slightly, they all ultimately pointed to the same landing page. 
Manual inspection revealed that the web page promotes a film associated with ``CandidateY". The landing page features a retro-styled embedded video player and a sign-up form requesting users’ names and email addresses in exchange for free access to the film. The page markets the film as available “for a limited time,” potentially encouraging urgency and engagement. This website was consistently linked across posts by accounts in the cluster.

Manual inspection further revealed that the content posted by these users was identical across accounts and consistently aimed at promoting ``CandidateY". Another notable pattern was the complete absence of hashtags—an uncommon trait that may reflect an intentional attempt to avoid categorization or detection by TikTok’s content moderation system. Finally, the posting behavior was tightly synchronized, with content appearing across accounts within five-minute intervals, further reinforcing evidence of coordinated activity.

For September, pruning the network yielded a dense cluster of 33 users, many of whom also appeared in the August cluster (see the \textit{Appendix}). These accounts exhibited nearly identical characteristics to those observed in August, but the number of such accounts had doubled. All usernames contained the string ``CandidateY'' and included the same URL in their video descriptions, linking to the external website shown in Figure~\ref{fig:co-domain_august}. As before, the posts lacked hashtags and displayed synchronized posting patterns (see the \textit{Appendix}). In October, we found no evidence of coordinated behavior using the co-domain behavioral trace, suggesting that this network either shifted its tactics or was dismantled. Notably, as of today, all accounts previously involved in promoting “CandidateY” have been removed or suspended from TikTok, potentially indicating moderation interventions.

\subsection{CIB Detection via Synchronous Posting}

\begin{figure*}[t]
\centering
\includegraphics[trim=0cm 0cm 0cm 1cm, clip=true, width=0.6\textwidth]{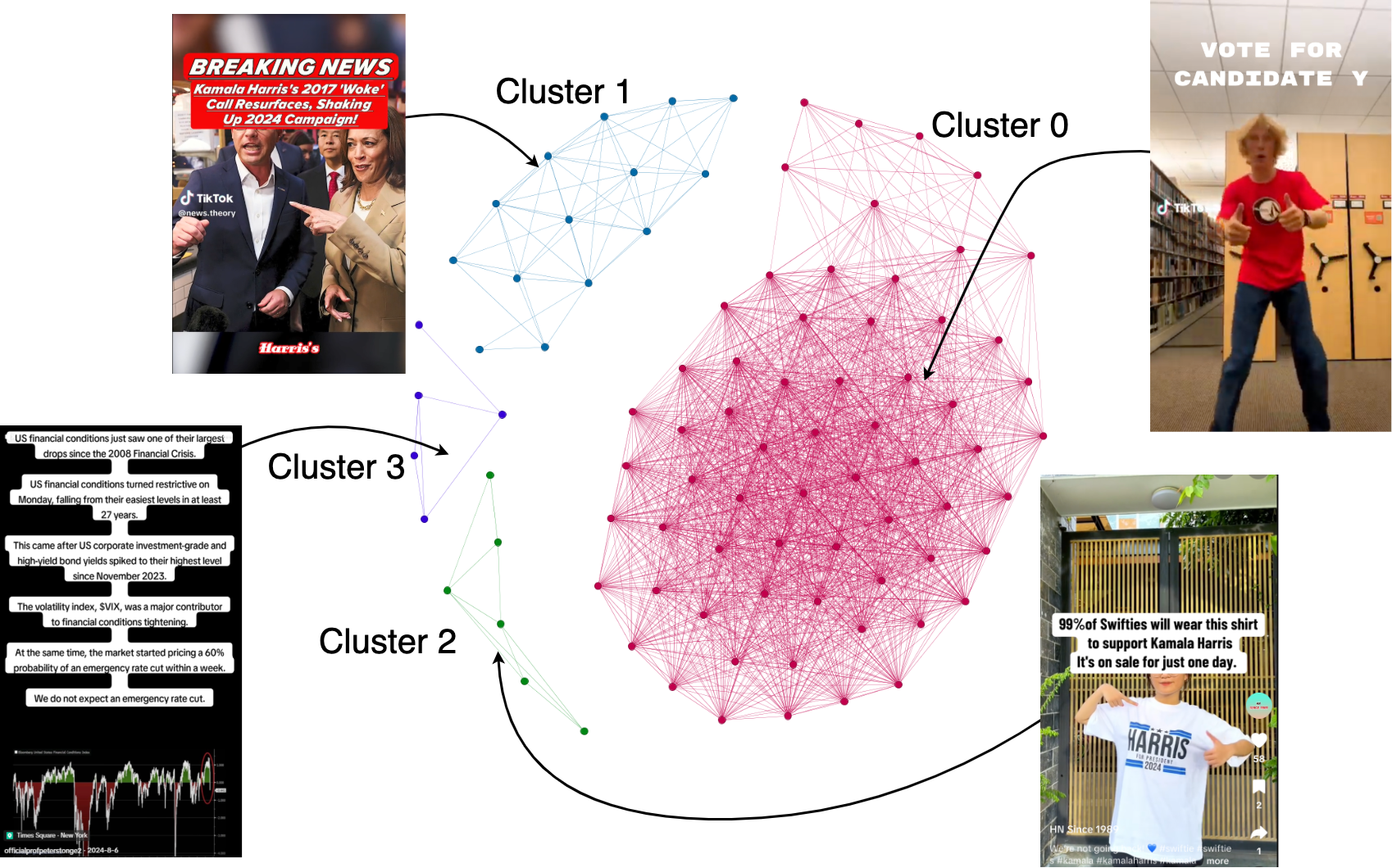} 
\caption{Clusters of coordinated accounts identified through the synchronous posting indicator.}
\label{fig:synchronized_communities_august}
\end{figure*}

To identify coordinated clusters with the synchronous posting behavioral trace, the network was pruned using a combination of edge filtering and node pruning \cite{cinus2025exposing}. First, edges with weights below the 99.5th percentile of similarity values were removed; then, node pruning was applied using a 95th percentile threshold, following the conservative criteria established in \cite{pacheco2021uncovering} and \cite{luceri2024unmasking}, respectively. For August, the pruning process resulted in four isolated clusters, as illustrated in Figure~\ref{fig:synchronized_communities_august}. Upon further investigation, all these clusters exhibited distinct forms of coordinated activity, which we detail as follows.

\textit{Cluster 0}: The largest cluster corresponds to the group of accounts promoting ``CandidateY", which we detected using the Co-Domain trace based on shared URLs in video descriptions. However, the time synchronization indicator analyzed here revealed an even broader set of accounts exhibiting similar characteristics, suggesting that not all users in this coordinated group relied on URLs to signal affiliation and promote content. All together, they consistently shared duplicate content within narrow time windows, indicating synchronized and likely centrally organized activity.

\textit{Cluster 1}: This cluster consists of accounts that have since been suspended from TikTok. Most usernames included the term “news” and exhibited structural similarities indicative of systematic coordination, e.g., the reuse of identical hashtag sequences across several posts. One account remained active during our analysis and was observed posting videos labeled “breaking news” along with snippets of political speeches. These consistent patterns in usernames, hashtags, and content suggest an orchestrated effort to drive political engagement by mimicking the style of credible news outlets and leveraging the visibility associated with breaking news. 

\textit{Cluster 2}:
Accounts in this cluster appear to be part of a coordinated marketing campaign. The same individual appears across all videos, and usernames follow a consistent pattern, each beginning with “hd.since”. These users posted duplicated video content promoting both Republican- and Democrat-themed merchandise, suggesting an effort to monetize rather than promote a specific ideological stance.

\textit{Cluster 3}:
This cluster is composed of accounts posting content related to the U.S. stock market. Identical financial videos were posted across multiple accounts (see the \textit{Appendix}). These accounts appeared to mimic or amplify the content of an official TikTok account with over 1 million likes and 140,000 followers, featuring usernames that included variations of the account's proper name and displaying similar profile photos. The observed behavior suggests an attempt to artificially boost the visibility of financial content through duplication and impersonation tactics.

Focusing on user activity in September, we identified two prominent clusters. The first appears to represent a group engaged in coordinated campaigns promoting alleged donation efforts for families in Gaza. Each user in this cluster typically posts videos calling for support or donations, often framed as personal or urgent appeals. A notable tactic employed by this group involves starting each video with an unrelated pre-clip, such as trending TikTok content, humorous moments featuring popular creators, or visually satisfying footage. This strategy likely serves to capture viewer attention, extend watch time, and maximize engagement before revealing the actual fundraising message. In addition, the group frequently leveraged election-related keywords in video descriptions to potentially hijack TikTok’s recommendation algorithm and broaden content visibility. Videos across this cluster shared identical hashtags and descriptions, but typically directed users to different PayPal donation links, often placed in user bios. In some cases, identical PayPal links were reused across multiple accounts. Further signals of coordination included the reuse of profile pictures and a consistent username pattern, with most usernames ending in the suffix ``family''. We also observed username changes and swapping among several accounts over time—a strategy previously linked to campaigns seeking to promote entities and run follow-back schemes \cite{pacheco2021uncovering, mariconti2017s}. Collectively, these patterns suggest a likely case of inauthentic coordination and raise concerns about deceptive campaign practices on the platform.

The second cluster comprises users whose usernames consistently begin with the name of ``CandidateX". These users regularly post TikTok videos that are highly similar in content and are synchronized in their timing, often appearing within minutes of each other. Notably, the videos shared by these users often consist of static images overlaid with political text, rather than original or dynamic audiovisual content. In many cases, the videos feature AI-generated voiceovers that narrate the on-screen text, contributing to a uniform and automated presentation style across accounts.
This lack of media diversity, combined with the repeated use of a fixed set of hashtags, reinforces the hypothesis that these users are engaged in a coordinated campaign aimed at promoting ``CandidateX". Interestingly, some of the users in this cluster also appear in the set of accounts identified through the hashtag sequence behavioral trace. While users identified through the hashtag sequence indicator tended to use a longer and more consistent hashtag order, users in this cluster employed a slightly shorter and reordered variant. Despite the variation in order, the core set of hashtags remained consistent, suggesting a possible effort to evade detection by subtly modifying posting patterns. Across this group, users demonstrated high similarity not only in hashtag usage but also in content structure, posting cadence, and strategic intent, all of which point to a deliberate, coordinated effort to promote support for ``CandidateX".

\begin{figure}[t]
\centering
\includegraphics[width=0.9\columnwidth]{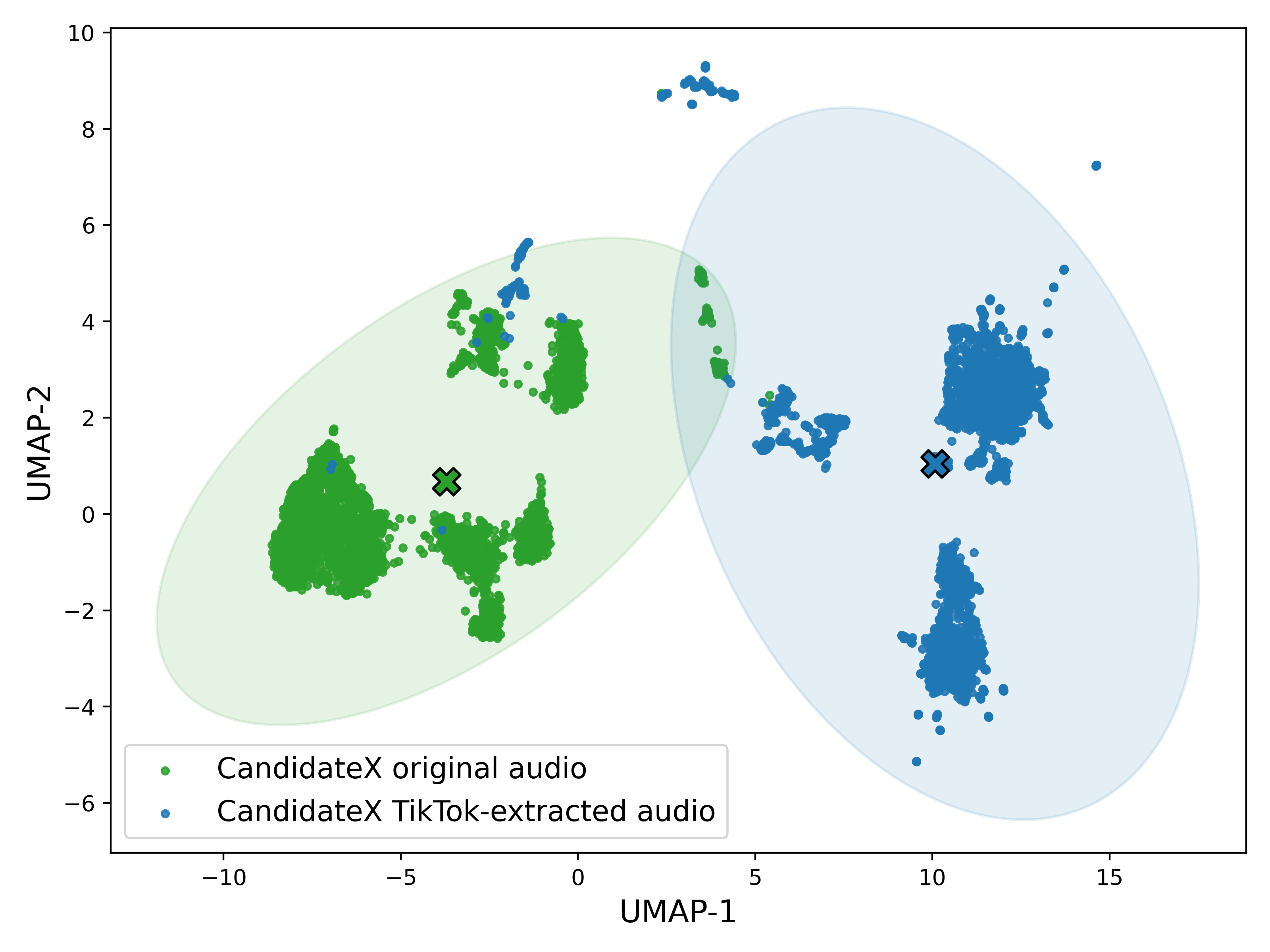} 
\caption{UMAP projection of audio embeddings: Each point corresponds to an individual audio clip, sourced either from verified, authentic recordings or from TikTok videos shared by coordinated accounts.}
\label{fig:candidatex_motivation_umap}
\vspace{-0.5cm}
\end{figure}

\subsection{CIB Detection via Speech Similarity}


For the month of August, the Speech Similarity network includes 102 accounts, all of which feature the string “CandidateY” in their usernames. This naming convention aligns with accounts previously identified through co-domain and synchronous posting traces. 
Although this coordinated network likely belongs to the broader campaign promoting “CandidateY”, its accounts and tactics, centered on repeatedly sharing videos with identical speech segments, differ significantly from those identified in previous analyses. 



The Speech Similarity network for September revealed a cluster of 42 accounts. These accounts posted videos with identical audio content at the exact same time, though each video featured distinct visuals. The content often promoted motivational messaging associated with ``CandidateX'', and account usernames followed consistent themes, combining ``CandidateX'' name with words like \textit{motivation, mindset, lesson}, or \textit{advices}. These usernames and the resulting network structure are shown in the \textit{Appendix}. Manual inspection reveals a recurring video format: each post begins with a photo of ``CandidateX'', followed by unrelated footage, such as scenic landscapes, inspirational imagery, or abstract visuals (examples in the \textit{Appendix}). Importantly, ``CandidateX'' appears only within the first five seconds of each video, suggesting a strategy of attention capture followed by narrative redirection, a tactic frequently employed in influence campaigns seeking to optimize user retention. 

In addition to visual consistency, a key commonality across these videos is the use of speech segments that resemble ``CandidateX'' voice. To assess whether these were authentic or synthetic, we conducted an audio-based forensic analysis using the ECAPA-TDNN model \cite{desplanques2020ecapa}, which is designed to capture subtle vocal cues and has been validated for detecting AI-generated or manipulated speech. We compared 426 suspect audio clips from TikTok videos shared by the coordinated network against 3,268 verified recordings of ``CandidateX'' from the \textit{In the Wild} dataset \cite{muller2022does}. All files were standardized, resampled to 16 kHz, and encoded via a pretrained ECAPA model to obtain audio embeddings. The resulting embeddings, visualized in Figure~\ref{fig:candidatex_motivation_umap} via UMAP dimensionality reduction, reveal a clear separation between verified and suspect audio samples (Fisher’s ratio $\approx$ 5.77). While a few TikTok clips may plausibly feature genuine speech, the overall distribution provides strong evidence that a substantial portion of the suspect audios were artificially synthesized. Manual annotations confirmed these results (see the \textit{Appendix} for more details).
This suggests a deliberate strategy to craft persuasive or emotionally resonant content, highlighting the evolving tactics of inauthentic campaigns that exploit the affordances of short-form video and synthetic media.  

The coordinated network resurfaced in October, alongside a separate spam campaign that hijacked election-related hashtags to promote a gaming application.

\begin{figure}[t]
\centering
\includegraphics[width=1\columnwidth]{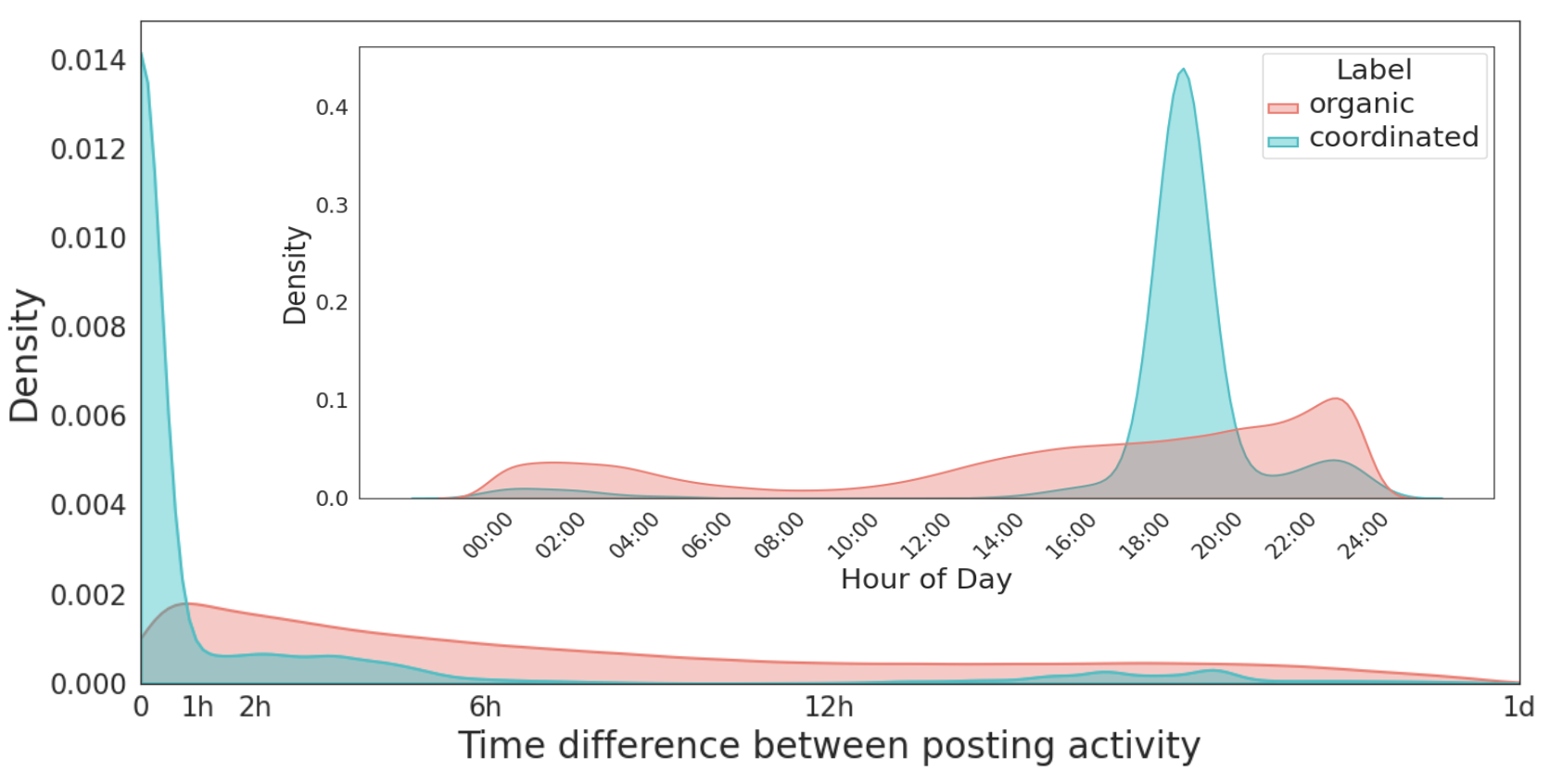} 
\caption{Probability density function of posting time differences and daily activity patterns of coordinated (video similarity network) vs. organic accounts}
\label{fig:candidatey_video_time_sync}
\vspace{-0.5cm}
\end{figure}

\subsection{CIB Detection via Video Similarity}

Through our video similarity framework, we uncovered a network of 67 coordinated accounts active in August. These accounts uniformly included the string “CandidateY” in their usernames, consistent with those identified through other behavioral traces such as speech similarity, synchronous posting, and co-domain. To quantify the overlap across these sets of coordinated accounts, we compute the Normalized Mutual Information (NMI) scores among the four networks. The consistently low values (see the \textit{Appendix} for details) confirm that, even within a single campaign, distinct sets of accounts engage in different coordination strategies. This aligns with prior research showing that influence campaigns often deploy multiple, minimally overlapping subnetworks to evade detection \cite{luceri2024unmasking,ng2022online}.

Figure~\ref{fig:candidatey_video_time_sync} visualizes the temporal signature of the coordinated activity of accounts in the video similarity network, comparing it against a baseline of organic (non-coordinated) accounts. The main panel displays the probability density of time gaps between their posts: coordinated accounts exhibit a sharp peak at 0 hours followed by a steep decline, while organic accounts show a broader, more gradual distribution. The inset further highlights the hourly activity patterns, revealing that coordinated accounts post almost exclusively between 6pm and 8pm UTC (2pm–5pm ET), in contrast to the more evenly distributed posting times of organic users, strengthening the evidence of inauthentic, synchronized amplification. The video similarity network for September and October continued to surface the same accounts uncovered in August. All the accounts in these clusters are no longer active on TikTok, possibly following moderation actions.


\subsection{Challenges of CIB Detection on TikTok}\label{sec:challenges}


TikTok’s Research API poses structural limitations, such as rate limits, incomplete metadata returns, and inconsistent quota fulfillment, that can hinder full behavioral coverage \cite{corso2024we,entrena2025tiktok}. To account for this, we simulated 5\% and 10\% data loss and found our detection methods remained robust, recovering nearly all coordinated accounts (see the \textit{Appendix} for more details). Mitigating these limitations at scale requires hybrid pipelines that combine the API with compliant third-party tools, as well as efficient sampling strategies and distributed crawling infrastructure.

While our framework uncovered several robust indicators of CIB, some behavioral traces did not reveal compelling evidence of coordination. These negative results are important for understanding the limitations and challenges associated with these alternative signals.

The analysis of \textit{Duet and Stitch} interactions revealed no suspicious clusters indicative of coordinated behavior, even after applying network pruning techniques based on eigenvector centrality and edge weight. This is likely because these features are organically used on TikTok to engage with popular content or express opinions on trending topics. In our dataset, users mainly performed Duet and Stitch to share their views on political discourse—a behavior more reflective of organic engagement than CIB. 

In examining \textit{co-domain} signals through user \textit{comments}, we found that the resulting network was sparse and fragmented. Most of the shared links pointed to other TikTok videos rather than external domains. This aligns with TikTok’s platform design, which de-emphasizes URL sharing and textual interaction. As a result, URL-based coordination through comments appears to be rare and did not provide useful insights into inauthentic behavior.

The \textit{co-reply} network yielded a cluster of highly active users engaged in political conversations, often replying to opposing viewpoints and defending specific candidates. While these users shared common hashtags and displayed strong ideological alignment, the lack of temporal synchronization, structural cohesion, or repetitive messaging suggests that their behavior was driven by organic partisanship rather than orchestrated coordination.


No substantial evidence of coordination emerged from textual similarity patterns in user \textit{comments}. Although users often expressed polarized opinions and used similar language, these similarities were more attributable to commenting on the same trending videos than to coordinated messaging. 
We also noted a high degree of overlap in the videos commented on by users in certain clusters, suggesting clustering was driven more by common exposure than by coordinated engagement.


\section{Discussion}
This study presents the first large-scale computational framework for detecting coordinated inauthentic behavior (CIB) on TikTok—a platform increasingly central to political discourse but largely understudied in the context of online manipulation. Building on prior work from platforms like Twitter/$\mathbb{X}$, we adapt and extend coordination detection methodologies to account for TikTok’s distinctive multimodal content and interaction mechanisms.

By leveraging diverse behavioral signals, including hashtag sequences, synchronized posting, shared web domains, and multimodal content reuse, we construct similarity networks that expose clusters of potentially coordinated users. Applied to a dataset centered on the 2024 U.S. Presidential Election, our approach reveals multiple instances of inauthentic activity, ranging from semi-automated political messaging campaigns to synthetic media usage and orchestrated content duplication. These signals, especially when converging across traces, indicate structured efforts to exploit TikTok’s algorithmic affordances to amplify political narratives.

Despite uncovering numerous examples of likely CIB, we find that many accounts involved in such behavior remain active on the platform. This highlights a potential gap between the detection of coordinated manipulation and enforcement actions, underscoring the need for proactive and scalable interventions tailored to TikTok’s ecosystem.

\paragraph{Key Takeaways.} Our analysis of TikTok-specific interaction patterns yields several actionable insights:

\textit{TikTok’s ecosystem fosters new and traditional modalities of coordination}: We observed both conventional strategies, like co-posting of URLs or copy-paste textual duplication, and TikTok-specific behaviors, such as manufactured split-screen videos and visuals featuring synthetic voiceovers. The cross-modal reuse of content (e.g., the same audio with different videos, and vice versa) suggests that TikTok’s algorithmic affordances are being actively gamed to amplify messaging while avoiding detection. Coordinated behaviors are increasingly adapted to the interaction norms and remix practices that define the platform’s user culture.

\textit{Multimodal coordination requires multimodal detection}: Many coordinated behaviors on TikTok—such as synchronized posting of near-identical videos or the use of synthetic voiceovers—would likely evade traditional, text-centric detection models. Unlike platforms where coordination often involves direct duplication of text or links, TikTok campaigns frequently rely on templated strategies and multimodal reuse: swapping visuals while retaining the same audio track, or employing split-screen formats where political content appears alongside unrelated, visually engaging footage. Detecting such tactics necessitates incorporating both visual and audio-based similarity metrics.

\textit{The growing role of Generative AI in CIB}: One of our key findings is the emergence of AI-generated voiceovers and synthetic speech used to simulate political figures and deliver persuasive messaging with minimal human labor. This trend marks a pivotal shift in coordination tactics: inauthentic campaigns are increasingly turning to generative AI to craft emotionally resonant, high-reach media at scale, sidestepping traditional, text-centric influence strategies.

\textit{CIB detection on TikTok is feasible, but coordination signals are highly fragmented}: Our framework successfully uncovered multiple CIB clusters linked to influence campaigns, yet each signal tended to expose largely disjoint sets of accounts. Even within the same campaign, different groups employed distinct tactics. This underscores that TikTok CIB does not coalesce around a single behavior, but instead relies on complementary strategies, requiring detection pipelines that integrate multiple weak signals.

\textit{Not all interaction features are useful}: Duets, stitches, and video-based replies did not produce strong signals of coordination. These are often used for reactive or oppositional content creation, not organized manipulation.

\subsubsection{Limitations.}
This study has several important limitations. First, while our method effectively identified coordination using several behavioral signals, others did not yield clusters indicative of inauthentic behavior. Future work is needed to assess whether this outcome reflects limitations in our current methods for capturing these signals, a lower prevalence of their deceptive use in CIB on TikTok, or characteristics specific to the dataset and discussion context analyzed. Second, we found that no single pruning or filtering strategy was consistently effective across all indicators, suggesting that the heterogeneity of user behavior on TikTok requires adaptive, trace-specific tuning. Third, our conservative approach prioritizes precision over recall, providing a lower bound on the prevalence of potential CIB-like activity. While this minimizes false positives, it may also overlook weaker or emerging coordination signals. 
Finally, while similar user clusters emerged across different indicators, they were not fully overlapping, suggesting that the signals capture complementary aspects of coordination. Future work will explore fusion strategies that integrate multiple indicators.

\subsubsection{Ethical Statement.}

We recognize both the positive and negative societal implications of CIB detection efforts. On the one hand, this research contributes to safeguarding the integrity of online discourse and offers computational tools to identify covert manipulation campaigns, potentially supporting democratic resilience. On the other hand, there are risks of misapplication, including the inadvertent labeling of organic, legitimate activism as inauthentic, or the weaponization of detection frameworks to suppress dissent. To mitigate these risks, we designed our framework to prioritize transparency and explainability, and we emphasize cautious interpretation of coordination signals.




\bibliography{aaai25}

\section{Appendix}

\subsection{Impact of Data Loss on Detection Strategies: A Robustness Assessment}
We assessed the robustness of our detection strategies under simulated data loss scenarios, randomly removing 5\% and 10\% of video data. 
Our method proved highly resilient, consistently retaining nearly all coordinated accounts under both data loss conditions.

\begin{itemize}
    \item \textbf{Hashtag Sequence:} Retained 98.5\% of coordinated accounts under both 5\% and 10\% data loss.
    \item \textbf{Co-Domain:} Retained 98.5\% of coordinated accounts under 5\% loss and 97.1\% under 10\% loss.
    \item \textbf{Speech Similarity:} Retained 98.0\% of coordinated accounts under 5\% loss and 91.2\% under 10\% loss.
    \item \textbf{Video Similarity:} Retained 97.1\% of coordinated accounts under 5\% loss and 92.6\% under 10\% loss.
\end{itemize}

\subsection{Manual Annotations Inter-Agreement}

Manual inspections were conducted by two annotators across several categories to validate the reliability of our findings. For annotations assessing whether usernames in distinct clusters contained references to either CandidateX or CandidateY, we observed perfect agreement (Cohen’s Kappa = 1.0). The same level of inter-annotator agreement was achieved when labeling: (i) synthetic vs. authentic voiceovers in CandidateX videos identified via the hashtag sequence indicator, (ii) Gaza-related accounts exhibiting handle rotation detected through synchronous posting, and (iii) manufactured split-screen versus duet videos.

Annotations of synthetic voices mimicking CandidateX (authentic vs. fake) yielded a Cohen’s Kappa score of 0.756, indicating substantial agreement. However, the strong class imbalance (most labels falling into the same category, i.e., fake) suppresses the Kappa value. The actual observed agreement was 421 out of 426 samples ($\sim$0.99), confirming that disagreement was extremely rare.


\begin{figure}[h!]
\centering
\includegraphics[width=1\columnwidth]{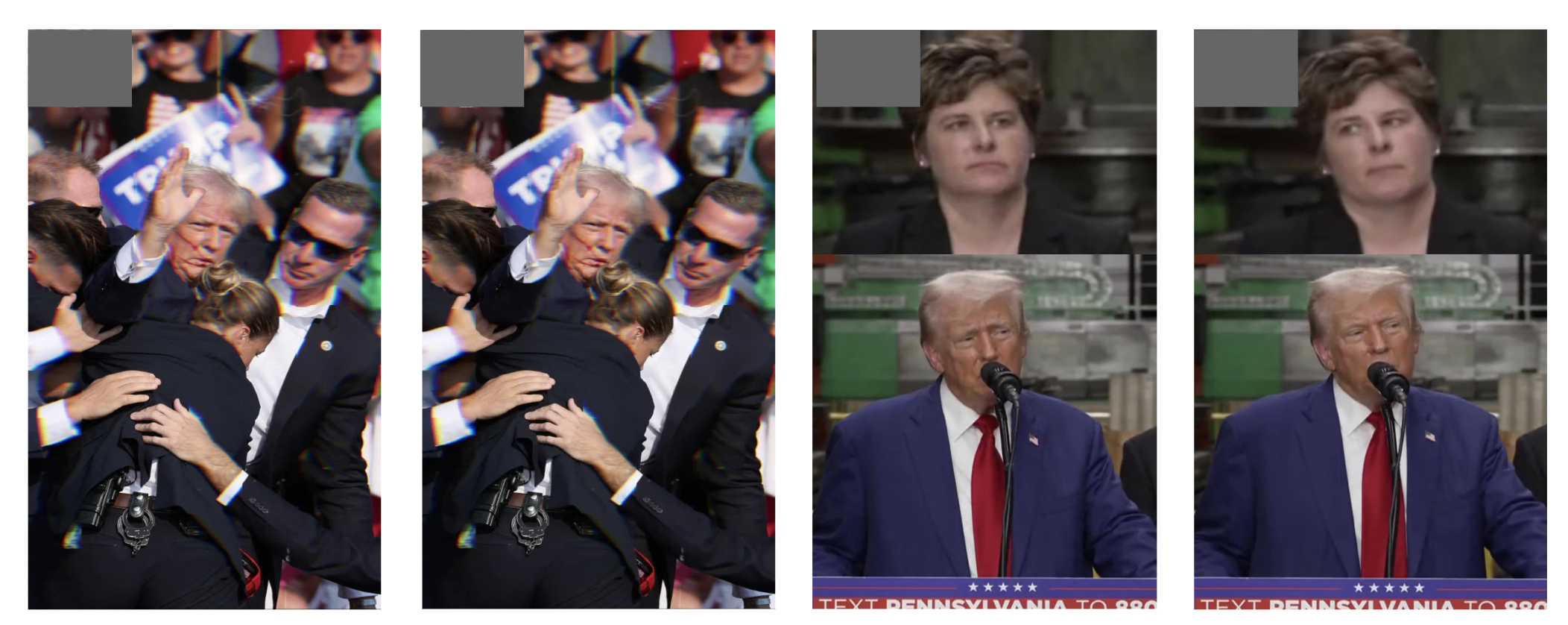} 
\caption{Examples of videos shared by coordinated accounts sharing the same hashtag sequences. The watermark in the top-left corner was
obscured to preserve anonymity and privacy.}
\label{fig:appendix}
\end{figure}

\begin{figure}[h!]
\centering
\includegraphics[width=1\columnwidth]{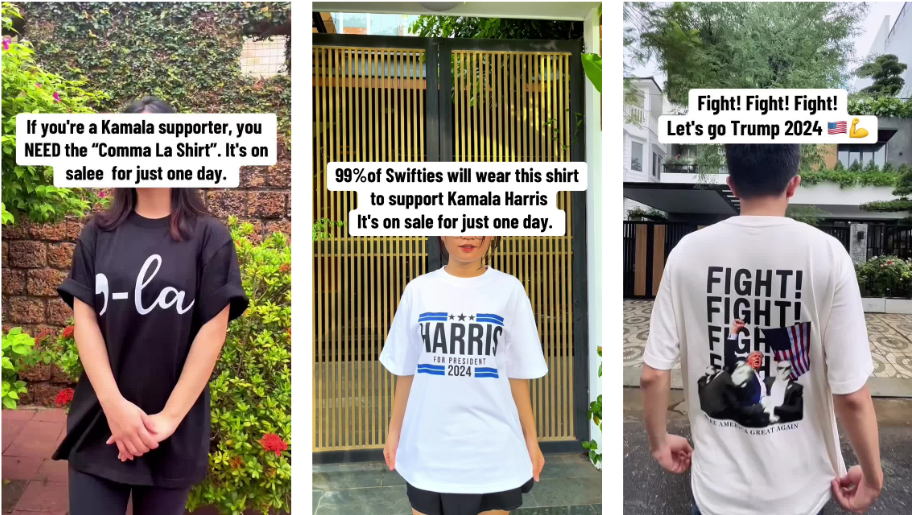} 
\caption{Video content shared by coordinated users in Cluster 2, identified through the synchronous posting indicator.}
\label{fig:cluster2}
\end{figure}

\begin{figure}[h!]
\centering
\includegraphics[width=1\columnwidth]{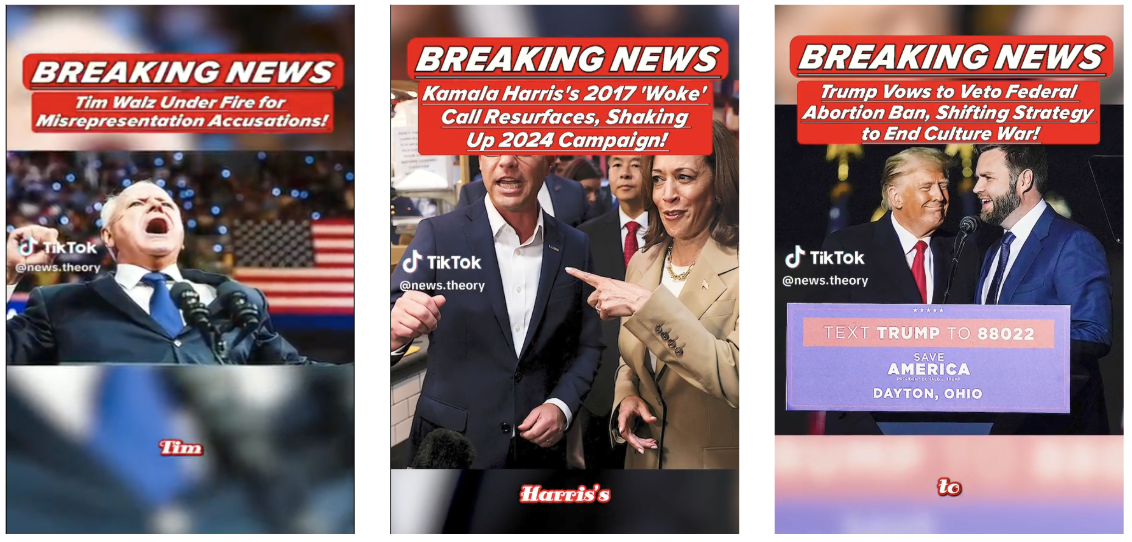} 
\caption{Video content shared by coordinated users in Cluster 1, identified through the synchronous posting indicator.}
\label{fig:cluster1}
\end{figure}

\begin{figure}[h!]
\centering
\includegraphics[width=1\columnwidth]{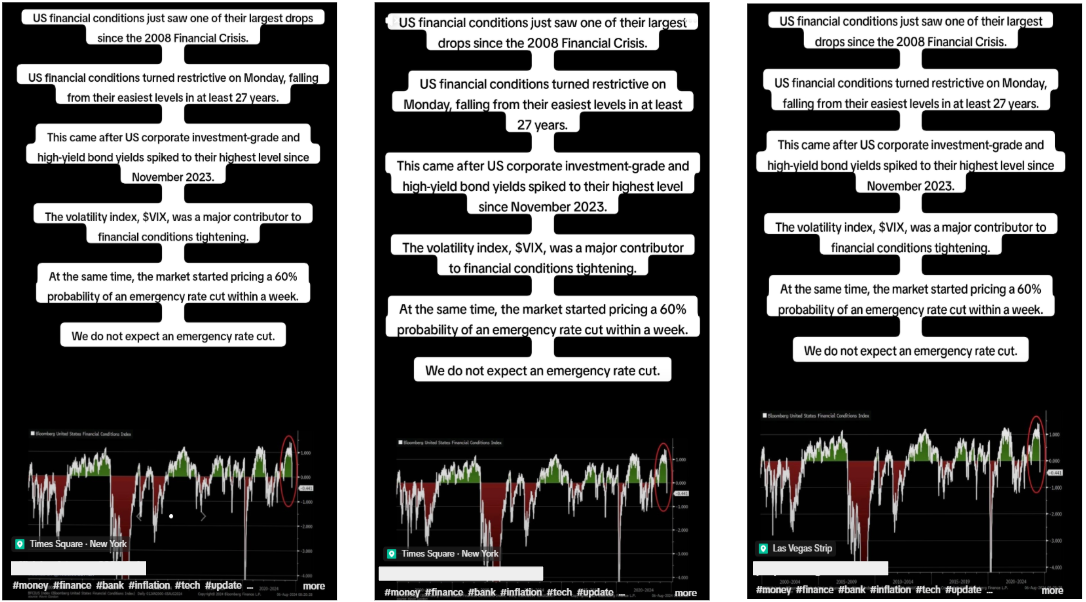} 
\caption{Video content shared by coordinated users in Cluster 3, identified through the synchronous posting indicator.}
\label{fig:cluster3}
\end{figure}

\begin{figure}[h!]
\centering
\includegraphics[trim=0cm 1cm 0cm 2cm, clip=true, width=0.99\columnwidth]{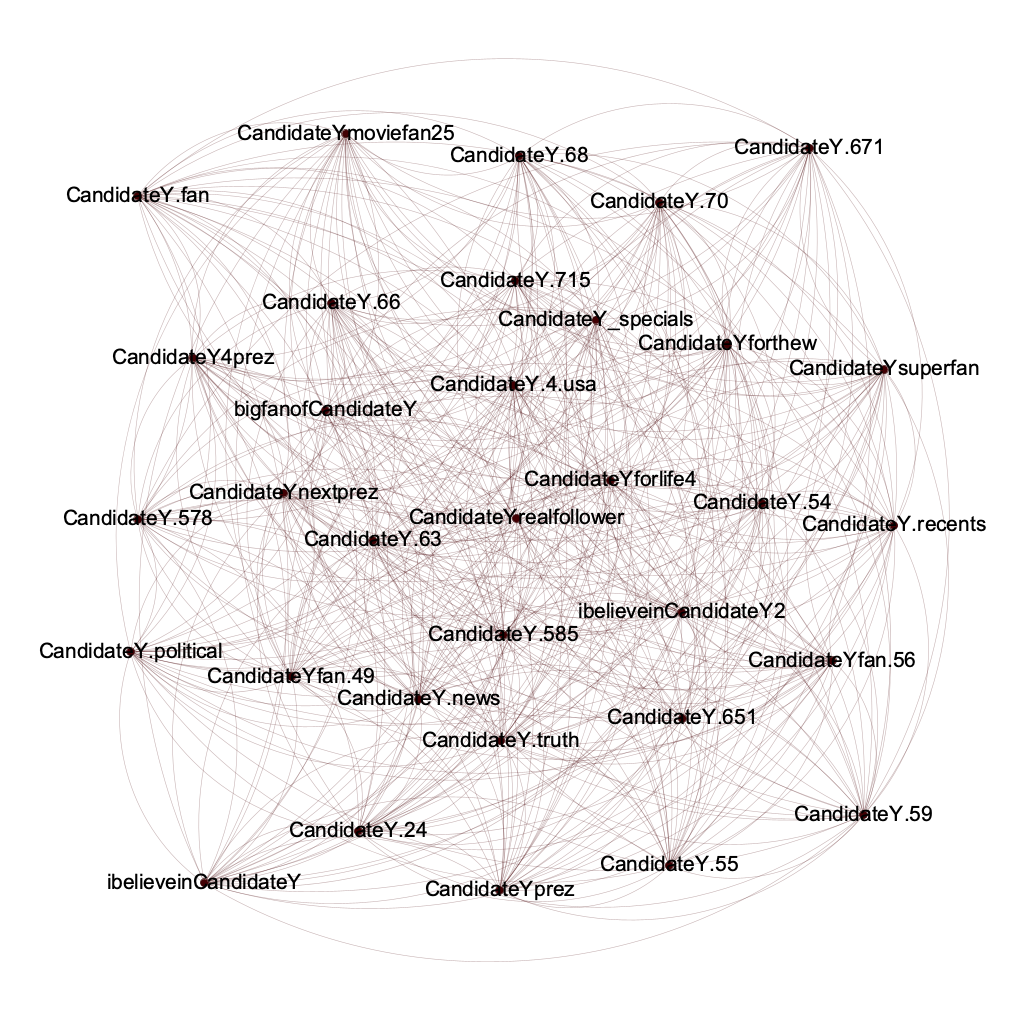} 
\caption{Dense cluster of coordinated accounts sharing the same web domains. Usernames across the cluster show strong similarity, all beginning with the same prefix.}
\label{fig:co-domain_sept}
\end{figure}

\begin{figure}[h!]
\centering
\includegraphics[width=1\columnwidth]{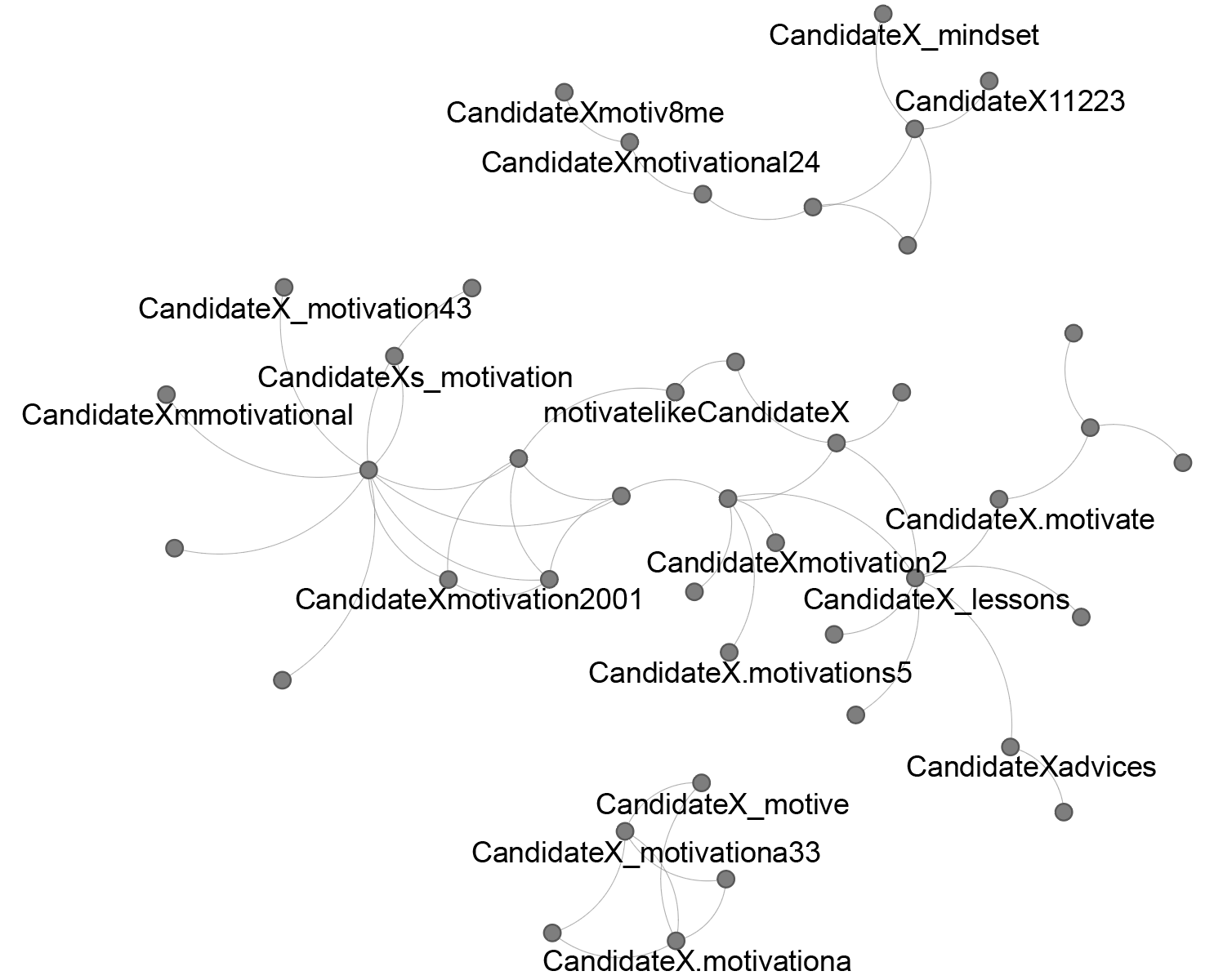} 
\caption{Coordinated cluster of accounts spreading ``CandidateX'' messaging. Usernames across the network display uniform prefixes and motivational themes.}
\label{fig:candidatex_motivation_network}
\end{figure}

\begin{figure*}[h!]
\centering
\includegraphics[width=0.75\textwidth]{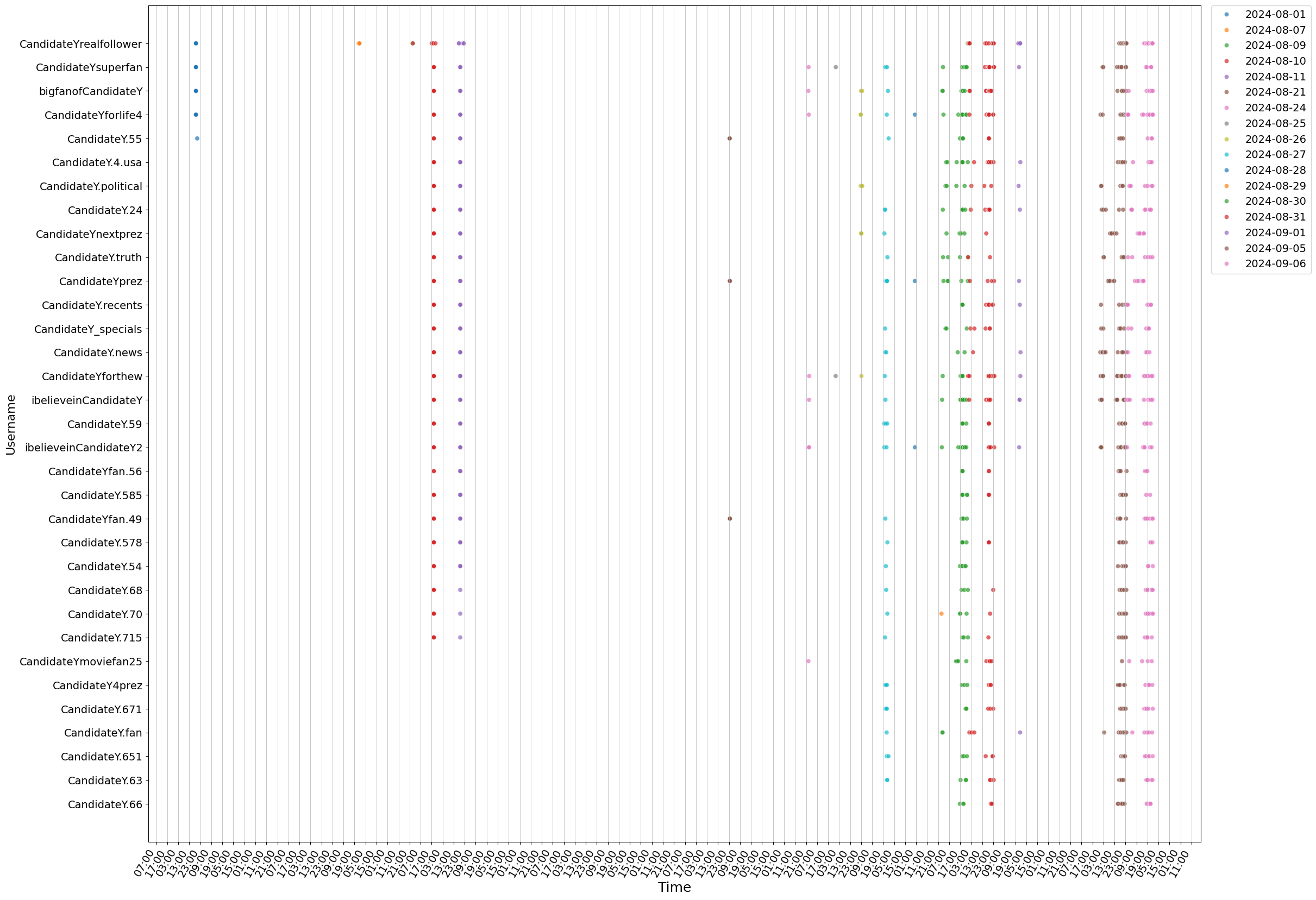} 
\caption{Synchronized behavior of coordinated accounts sharing the same web domains promoting ``CandidateY".}
\label{fig:sync_codomain_candidatey}
\end{figure*}

\begin{figure*}[h!]
\centering
\includegraphics[clip=true, width=0.8\textwidth]{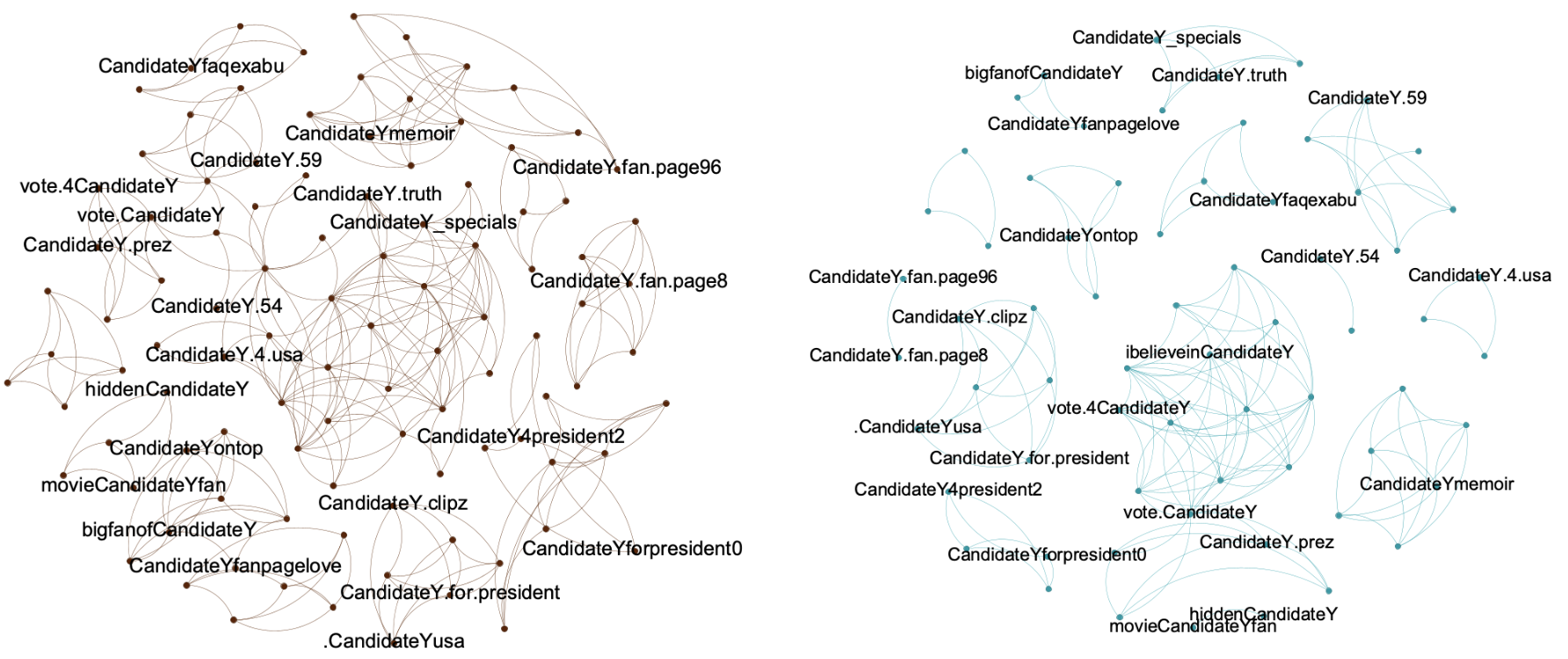} 
\caption{Cluster of coordinated accounts sharing near-identical video transcripts (left panel) and visuals (right panel). Usernames within both clusters show strong similarity, all beginning with a prefix linked to a U.S. presidential candidate, denoted here as “CandidateY” to pre-
serve anonymity and privacy. To reduce visual clutter, we
display only the usernames of accounts that posted more
than ten videos. Despite sharing similar usernames, the two networks exhibit minimal overlap (NMI = 0.1).
}
\label{fig:candidatey_text_sim_video_sim_network}
\end{figure*}

\paragraph{Keywords and Hashtags used in the Data Collection}
\begin{itemize}
    \item election2024, Election2024
    \item US Elections, USElections, us elections, uselections
    \item 2024Elections, 2024 Elections, 2024 elections, 2024elections, 2024election
    \item 2024PresidentialElections, 2024 Presidential Elections, 2024presidentialelections, 2024 presidential elections
    \item saveamerica2024
    \item presidentbiden, Biden, biden, bidenharris, JoeBiden, Joe Biden, joebiden, joe biden, joseph biden, Joseph Biden
    \item Biden2024, biden2024, bidenharris2024
    \item Donald Trump, donald trump, DonaldTrump, donaldtrump, donaldtrump2024
    \item Trump2024, trump2024, trumpsupporters, trumptrain
    \item republicansoftiktok, conservative
    \item MAGA, maga, makeamericagreatagain, ultramaga, KAG
    \item Republican, presidenttrump, trumpismypresident, letsgobrandon
    \item GOP, CPAC
    \item NikkiHaley, nikkihaley
    \item DeSantis, RonDeSantis, desantis, rondesantis
    \item RNC
    \item democratsoftiktok, democratsarehot, thedemocrats
    \item voteblue2024, vote blue
    \item DNC, dnc
    \item kamalaharris, kamala harris
    \item mariannewilliamson, williamson2024
    \item deanphillips, phillips2024
    \item democratic party, Democratic party
    \item republican party, Republican party
    \item Third party, third party
    \item Green party, green party
    \item Independent party, independent party
    \item RFKJr, RFK Jr., RFK Jr, rfkjr, rfkjr., rfk, Robert F. Kennedy Jr., robert f. kennedy Jr.
    \item jill stein, jillstein, Jill Stein, JillStein
    \item CornellWest, cornellwest
\end{itemize}

\begin{figure}[t]
\centering
\includegraphics[width=0.85\columnwidth]{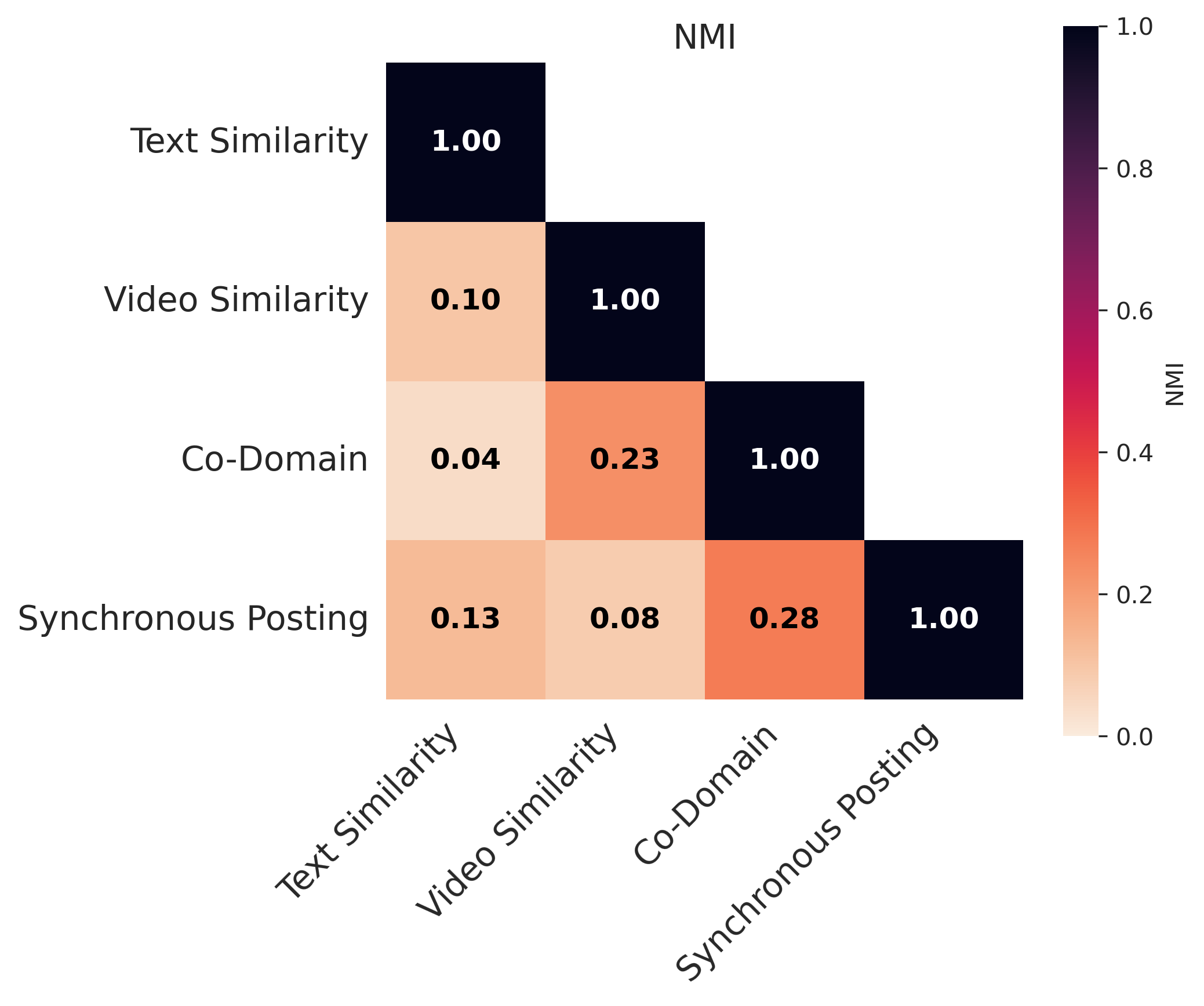} 
\caption{NMI score for CandidateY accounts detected across multiple strategies}
\label{fig:NMI}
\end{figure}
\vspace{-2em} %
\begin{figure}[h!]
\centering
\includegraphics[width=1\columnwidth]{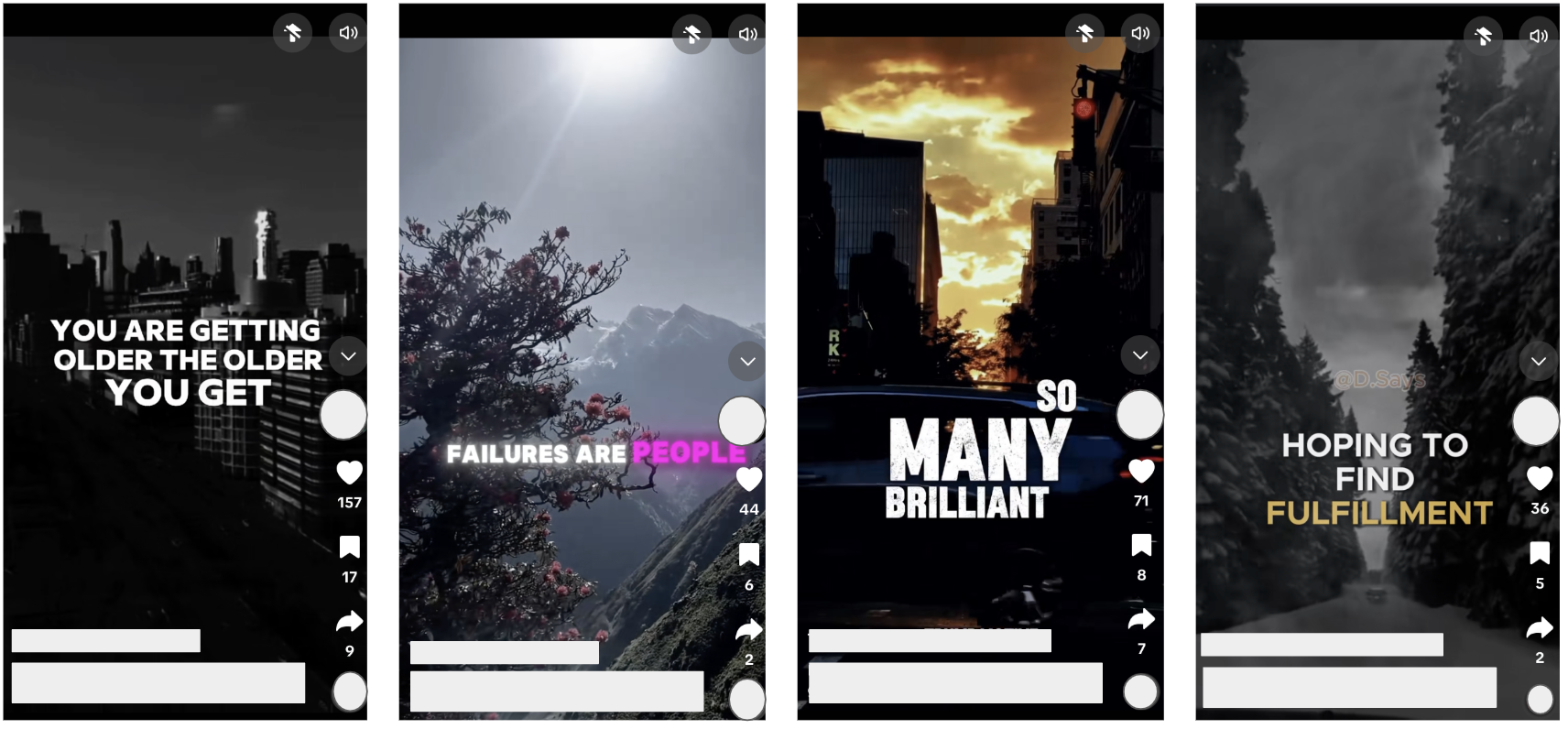} 
\caption{Examples of TikTok videos from a coordinated network employing synthetic ``CandidateX'' voiceovers to deliver motivational messages.}
\label{fig:candidatex_motivation_example}
\end{figure}

\end{document}